\documentclass[superscriptaddress, preprintnumbers,amsmath,amssymb]{revtex4}
\usepackage{graphicx}
\usepackage{dcolumn}
\usepackage{bm}
\usepackage{latexsym}
\begin{document}
\title{{\bf 
Distribution of lifetimes of kinetochore-microtubule attachments:\\interplay of energy landscape, molecular motors and microtubule (de-)polymerization}{\footnote{The first two authors contributed equally}}}
\author{Ajeet K. Sharma}
\affiliation{Department of Physics, Indian Institute of Technology Kanpur, 208016}
\author{Blerta Shtylla}
\altaffiliation[Current Address:]{ Department of Mathematics, Pomona College, Claremont, CA 91711}
\affiliation{ Mathematical Biosciences Institute, The Ohio State University, 1735 Neil Avenue, Columbus, OH 43210, U.S.A.}
\author{Debashish Chowdhury}
\affiliation{Department of Physics, Indian Institute of Technology Kanpur, 208016, and\\
Max-Planck Institute for Physics of Complex Systems, 01187 Dresden, Germany}
\begin{abstract}
Before a cell divides into two daughter cells, chromosomes  
are replicated resulting in two sister chromosomes embracing 
each other. Each sister chromosome is bound to a separate proteinous 
structure, called kinetochore (kt), that captures the tip of a filamentous 
protein, called microtubule (MT). Two oppositely oriented MTs pull the 
two kts attached to two sister chromosomes thereby pulling the two sisters 
away from each other. Here we theoretically study an even simpler system, 
namely an isolated kt coupled to a single MT; this system mimics an 
{\it in-vitro} experiment where a single kt-MT attachment is reconstituted 
using purified extracts from budding yeast. Our models not only account 
for the experimentally observed ``catch-bond-like'' behavior of the kt-MT 
coupling, but also make new predictions on the probability distribution of 
the lifetimes of the attachments. In principle, our new predictions can be 
tested by analyzing the data collected in the {\it in-vitro} experiments 
provided the experiment is repeated sufficiently large number of times. 
Our theory provides a deep insight into the effects of (a) size, (b) 
energetics, and (c) stochastic kinetics of the kt-MT coupling on the 
distribution of the lifetimes of these attachments.

\end{abstract}

\maketitle

\section{Introduction}

Chromosomes, the genetic material of a cell, are duplicated and properly 
segregated before the cell divides into two daughter cells \cite{lodish}. 
Each of the sister chromatids, that result from chromosome replication, 
is bound to a proteinous structure, called kinetochore (kt) that, in turn, 
is coupled to the plus ends of stiff polar filaments called microtubules 
(MT) \cite{cheeseman08}. The negative ends of these MTs are located at 
the poles of the fusiform structure, called mitotic spindle \cite{karsenti01}. 
The process of chromosome segregation, called mitosis \cite{mcintosh12}, 
is carried out in eukaryotic cells by the dynamic mitotic spindle which 
is self-organized from its components for this purpose. There are strong 
indications that molecular motors \cite{chowdhury13a,chowdhury13b} are 
located at the 
kt-MT interface (though not in all eukaryotes); these motor proteins, 
which are capable of generating force by hydrolyzing ATP, are believed 
to generate poleward force or anti-poleward force depending on the family 
to which they belong. However, polymerizing and depolymerizing MTs are 
also found to make significant  contributions to the forces that cause 
chromosomal movements 
\cite{joglekar02,gardner05,civelekoglu06,mcintosh12}. 

The force-generation capability of depolymerizing MTs have been 
demonstrated {\it in-vitro} \cite{peskin95}. 
However, how a MT maintains contact with the kinetochore while 
depolymerizing from its tip and how it exerts force on the kt 
remain challenging open questions in spite of the recent progress 
in identifying the molecular components of a kt and their spatial 
organization 
\cite{cheeseman08}, 
Moreover, in most of the eukaryotic organisms multiple MTs couple to 
each kt; how the (de-)polymerization of mutiple MTs are coordinated 
is still a mystery. {\it Budding yeast} is a much simpler system 
because each kt is coupled to only a single MT. But, even in 
budding yeast, the two embraced sister chromatids are coupled to 
two MTs approaching them from the two opposite poles of the spindle. 

In this paper we theoretically study an even simpler system, namely 
an isolated kt coupled to a single MT; this system mimics 
an {\it in-vitro} experiment \cite{akiyoshi10} where a single kt-MT 
attachment was reconstituted using purified extracts from budding yeast.
In those {\it in-vitro} experiments the average lifetime was found 
to vary nonmonotonically with the externally applied load tension. 
In other words, the kt-MT attachment is stabilized (instead of getting 
destabilized) at sufficiently small load tension. Similar tension-induced 
stabilisation of attachment between E-coli fimbriae and target cells 
have been interpreted in the past in terms of the concept of ``catch-bond'' 
\cite{thomas08a}. Akiyoshi et al. \cite{akiyoshi10} developed a 
2-state kinetic model that involves four phenomenological rate constants. 
By assuming the rate constants to vary exponentially with the load tension, 
they obtained a reasonably good fit of the model predictions with their 
experimental data.

Here we do not attempt any quantitative comparison with the experimental 
data reported in ref.\cite{akiyoshi10}. Instead, our approach 
allows to derive {\it analytical} expressions for the lifetimes of the 
kt-MT attachments within the framework of the simple theoretical 
models that we develop here. These models describe the polymerization- 
depolymerization of the MT explicitly and capture the effects of 
its interaction with the coupler by a potential energy landscape 
that gets modified when the MT is subjected to an external tension. 
Therefore, our results for the mean lifetimes provide deep physical 
insight into the interplay of opposing forces and competing kinetic 
processes that, together, determine the stability of the kt-MT coupler. 
We also make theoretical predictions on the probability distribution 
of the lifetimes which also indicate the probability of survival of an 
attachment at least for a time interval $t$ after the attachment is 
established. In principle, our new predictions can be tested by analyzing
the data collected in the {\it in-vitro} experiments provided the experiment
is repeated sufficiently large number of times.

This paper is organized as follows. In sections \ref{sec-minimalcont} 
and \ref{sec-minimaldisc} we develop a {\it minimal theoretical model} 
of a device that couples a single MT with a single kt and subject it 
to an external tension. Space is represented by a continuous variable in 
section \ref{sec-minimalcont} and by a discrete variable in section 
\ref{sec-minimaldisc}. Our results on the distribution of the lifetimes 
of the model kt-MT attachments, derived analytically in section 
\ref{sec-minimalcont}, are compared with the corresponding numerical 
data, obtained by numerical simulations, in section \ref{sec-minimaldisc}. 
Next, in section \ref{sec-realistic}, we extend the minimal model by 
representing the kt-MT interaction by a more realistic potential. 
Calculation of the full distribution of the lifetimes in this case is 
too difficult to be carried out analytically. Therefore, in this case 
we have directly calculated the mean lifetime of the kt-MT attachments 
and found that the realistic potential leads to quantitative changes 
without affecting the qualitative features of the kt-MT lifetimes observed 
in the  minimal model. Finally, motivated by recent evidence that the 
kt-MT coupler in mammalian cells might be a hybrid nano-device, composed 
of spatially separated active and passive components, we have extended our 
study also to a simple model of a hybrid coupler in appendix \ref{sec-hybrid}.

\section{A minimal model: continuum formulation and results}
\label{sec-minimalcont}

In the first version of the minimal model space is 
represented by a continuous variable and the kinetics of the system is 
formulated in terms of a Fokker-Planck equation. We find it more convenient 
to derive our analytical results using this formulation. For the convenience 
of numerical simulations, in the next section, we discretize the same 
model following prescriptions proposed earlier by Wang, Peskin and Elston 
(WPE) \cite{wang03,wang07}. We also compare the results of the two versions.

\begin{figure}[tb]
\begin{center}
\includegraphics[width=0.85\columnwidth]{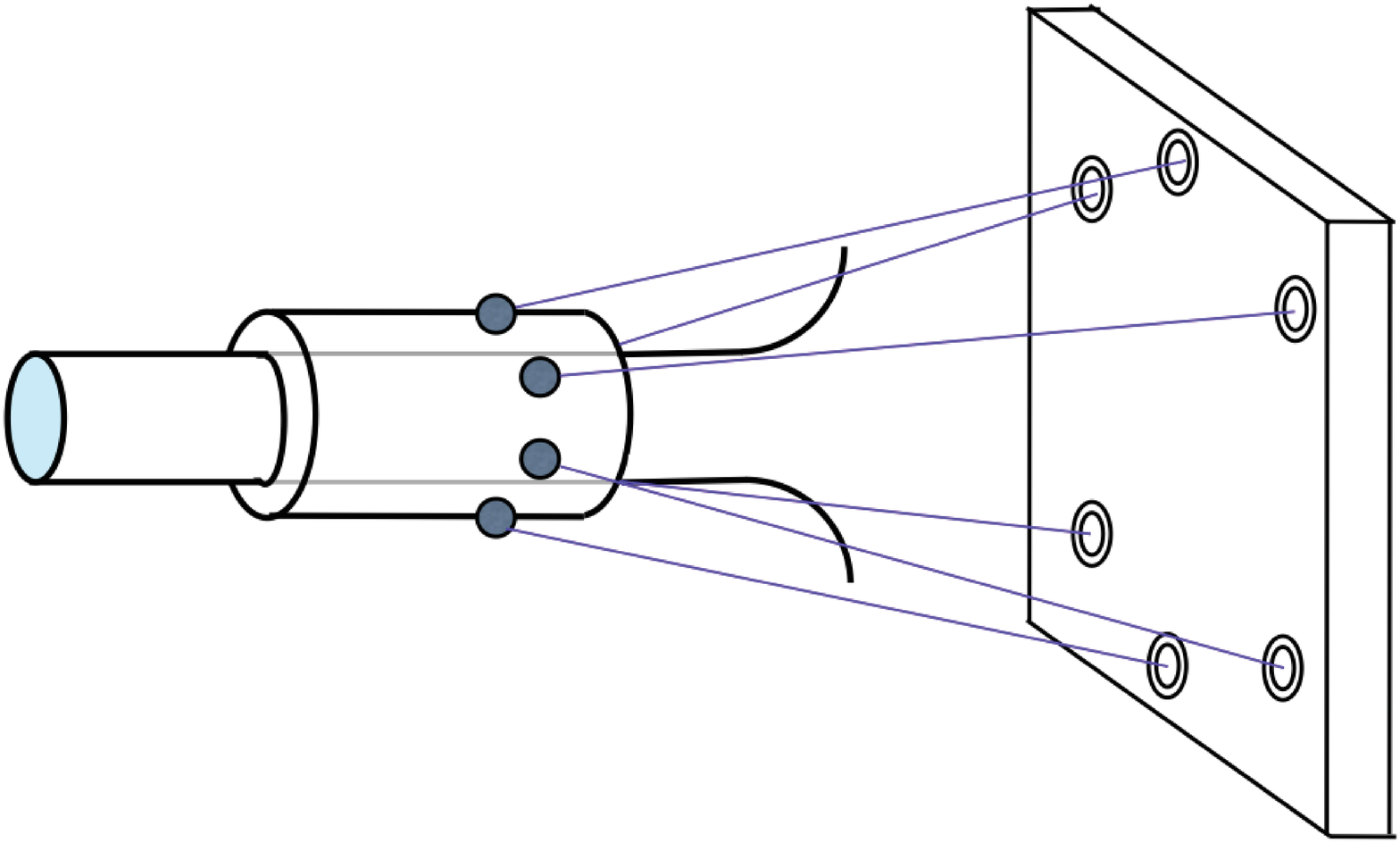}
\end{center}
\caption{(Color online) A schematic depiction of the minimal model. The sleeve-like coupler is rigidly connected to the kinetochore plate that is represented by a rectangular surface. The plus-end of the MT is closer to the kinetochore plate while the distal end is the minus end of the MT.
}
\label{fig-toymodel}
\end{figure}

For the minimal model we do not need to make any assumption about 
the molecular constituents and their spatial organization in the 
coupler. The only postulates are as follows:\\ 
{\it Postulate (a)}: the energy of the 
system is lowered monotonically with the increasing overlap between 
the inner surface of the cylindrically shaped coupler and the outer 
surface of the MT, and the corresponding binding energy is proportional 
to the projected length of this overlap along the MT axis;\\ 
{\it Postulate (b)}: the 
rate of depolymerization of a MT is suppressed by external force. 

To our knowledge, the postulate (a) is consistent with the scenario 
envisaged originally by Hill \cite{hill85}, except for the fact 
that the ``roughness'' of the MT-coupler interface in the Hill model,  
which is also an integral part of our realistic model (described in the 
next section), is not captured by the minimal version of our model.
The existence of the sleeve was postulated by Hill \cite{hill85} long 
before information on the the molecular structure of MT-kinetochore 
coupler began to emerge from experiments. The current knowledge on 
the inventory of the mitotic machinery have identified the plausible 
candidates that give rise to the effectively sleeve-like coupler. 
The Dam1 ring (also called DASH) and the Ndc80 complex seem to be the 
strongest candidates for the components of the coupler 
\cite{westermann07,asbury11,buttrick11,foley13}.

The postulate (b) is supported by the observation of Franck et al. 
\cite{franck07} in their {\it in-vitro} experiments that the rate 
of depolymerization of MT is suppressed by externally applied tension. 
It has been known from quite some time that the tips of the depolymerizing 
protofilaments are curled radially outward from the central axis of 
the cylindrically shaped MT. Based on this fact, Franck et al.\cite{franck07} 
speculated that the Dam1 ring complex may transmit the external tension 
to the curled tips of the protofilaments thereby tending to straighten 
them. Such straightening of the protofilaments is likely to suppress 
the tendency of the protofilaments to peel away from the depolymerizing 
tip of the MT. 
It is worth emphasizing that our results presented here require only 
the validity of the postulates (a) and (b) irrespective of the nature of 
the underlying cause of their validity.

\subsection{Continuum formulation of the minimal model}

The length of the ``coupler''is denoted by $L$. In the continuum 
formulation of the stochastic kinetics, the time-dependent variable 
$x(t)$ denotes the instantaneous length of overlap between the outer 
surface of the MT and inner surface of the coaxial cylindrical coupler. 
Thus, $x=0$ and $x=L$ corresponds to minimum and maximum overlap, 
respectively; the MT is just on the verge of exiting the coupler when 
$x=0$. 
 
Guided by our postulate (a), in the minimal version of our model we 
capture the MT-coupler interaction by a potential energy that is 
proportional to the overlap $x$, i.e.,    
\begin{equation}
U_{b}(x)= - B x.
\end{equation}
parametrized by $BL$, the depth of the potential at $x=L$. The form 
of this interaction in the more realistic version of our model is 
given in section \ref{sec-realistic}.

\begin{figure}[tb]
\begin{center}
\includegraphics[width=0.9\columnwidth]{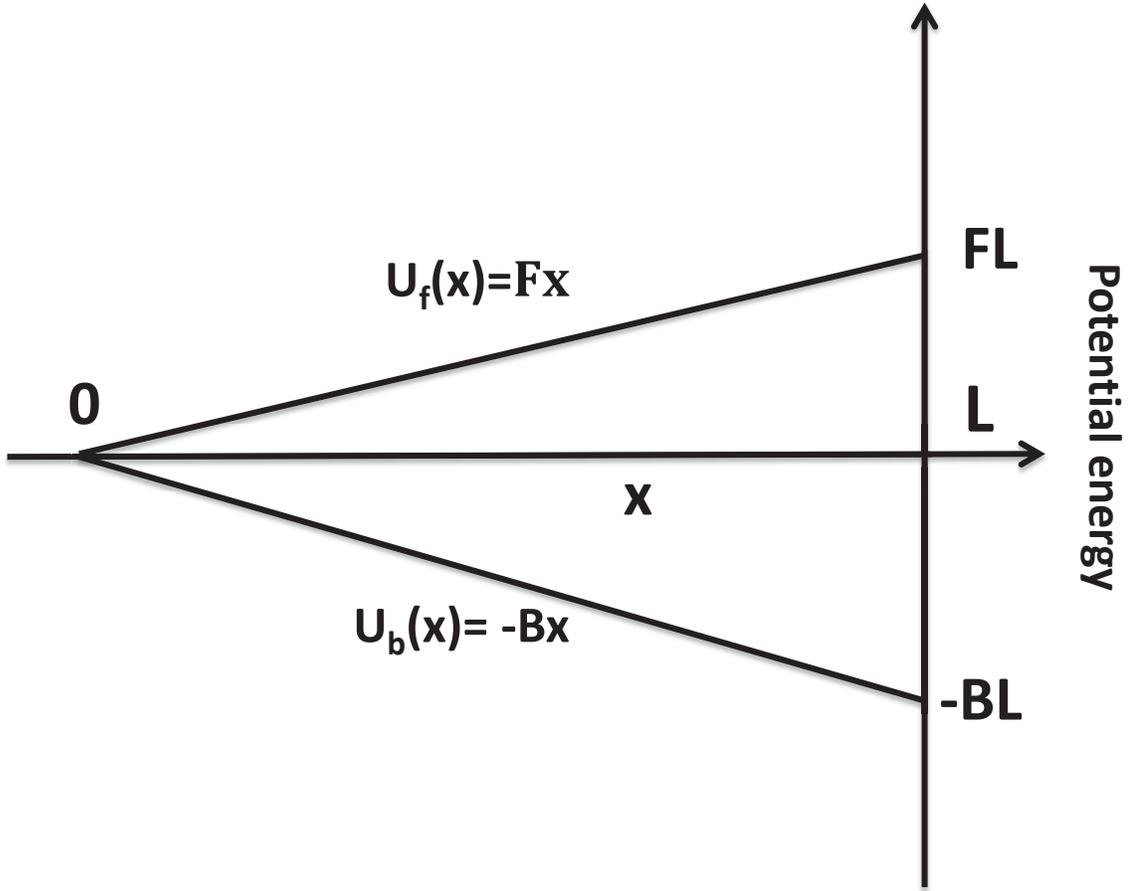}
\end{center}
\caption{Hypothesized potential $U_{b}(x)$ and $U_{f}(x)$ are plotted 
against the instantaneous length of overlap $x(t)$ between the outer 
surface of the MT and inner surface of the coaxial cylindrical coupler. 
}
\label{potential}
\end{figure}

Similarly if the MT is pulled outward away from the coupler by a 
constant external force $F$ (also referred to as a load tension), 
then this force can be derived from the corresponding potential,
\begin{equation}
U_{f}(x)= F x
\end{equation} 
$U_{b}(x)$ and $U_{f}(x)$ are plotted in fig. (\ref{potential}).
Net potential felt by the kinetochore is 
\begin{equation}
U(x)=U_{b}(x)+U_{f}(x).
\end{equation}

We use the symbols $\alpha$ and $\beta$ to denote the addition and removal 
rates of the tubulin subunit from the MT. 
When subjected to a load tension, according to our postulate (b), the 
depolymerization rate $\beta(F)$ of the MT decreases with increasing 
strength of the tension $F$ \cite{franck07}. More specifically, we assume  
an exponential decrease of $\beta(F)$ such that 
\begin{equation}
\beta(F)=\beta_{\text{max}}~ e^{-F/F_{*}}
\label{eq-Fdepbeta}
\end{equation} 
where $\beta_{\text{max}} \equiv \beta(0)$ is the depolymerization rate in the 
absence of any external force and $F_{*}$ is the characteristic load 
force at which the MT depolymerization rate is an exponentially small 
fraction of (i.e., a factor of $1/e$ smaller than) $\beta_{\text{max}}$. 
Our motivation for assuming the specific 
exponential form (\ref{eq-Fdepbeta}) and the key role of the characteristic 
force $F_{*}$ will be discussed later in this paper.

Stochastic movement of the system can be described by a {\it hypothetical 
diffusing particle in an external potential U(x)}. The drift caused by 
the addition and removal of the tubulin from MT give rise to an 
additional term $(\alpha-\beta)\ell$ to the drift velocity. 
Note that the position of the hypothetical particle (i.e., the overlap 
$x$) can change by any of the following processes:\\
(i) Even in the absence of any external force, the MT has a natural 
tendency to approach $x=L$ because the system can lower its energy 
by increasing the overlap between the outer surface of the MT and 
the inner surface of the coupler. In other words, the potential 
$U_{b}(x)$ gives rise an effective force that spontaneously pulls 
the MT into the coupler.\\
(ii) The external load tension can pull the MT outward with respect to 
the coupler thereby decreasing the overlap $x$. 
Alternatively, an external force can push the MT into the coupler as 
long as $x$ remains non-zero. \\
(iii) The overlap $x$ can decrease, as long as it is non-zero, because 
of the depolymerization of the MT. \\
(iv) The overlap $x$ can increase because of the polymerization of MT. \\
(v) Finally, even in the absence of polymerization or depolymerization 
and force-induced movements, the position $x$ can change because of 
its one-dimensional diffusion; $D$ being the corresponding diffusion constant.

Therefore, in the overdamped limit,  
\begin{equation}
\frac{dx}{dt}=-\frac{1}{\Gamma}\frac{dU(x)}{dx} + (\alpha-\beta)\ell + \frac{\eta(t)}{\Gamma}  
\label{eq-langevin}
\end{equation}
where $\Gamma$ is the phenomenological coefficient of the viscous drag and 
$\eta(t)$ is a Gaussian white noise.

Let $P(x,t)$ denote the probability of finding the overlap $x$ at time $t$.
The Fokker Planck equation \cite{balakrishanan},
\begin{equation}
\frac{\partial P(x,t)}{\partial t}= D \frac{\partial^2 P(x,t)}{\partial x^2}-c\frac{\partial P(x,t)}{\partial x},
\label{main}
\end{equation}
that corresponds to the stochastic differential equation (\ref{eq-langevin}),  
is essentially a diffusion equation with an additional drift term
where the time-independent, but tension-dependent, net drift velocity is
\begin{equation}
c=\underbrace{\frac{B-F}{\Gamma}}_\text{drift due to two forces}+\underbrace{(\alpha-\beta_{\text{max}}~ e^{-F/F_{*}})\ell}_\text{drift due to polymerization kinetics}
\label{myc}
\end{equation}
and D is the diffusion constant. The interplay of the 
four key components are shown clearly in eqn.(\ref{myc}). The external 
load tension $F$ competes against $B$ while MT depolymerization 
competes against polymerization. The load tension $F$ has two mutually 
opposite effects on the kinetics: (i) it directly tends to decrease $x$ 
by pulling out the MT from the coupler, and (ii) it indirectly assists 
increase of $x$ by suppressing $\beta(F)$ that competes against $\alpha$.

The equation (\ref{main}) can also be written as an equation of continuity
\begin{equation}
\frac{\partial P(x,t)}{\partial t} + \frac{\partial J(x,t)}{\partial x} = 0
\end{equation}
for the probability density $P(x,t)$, with the probability current density 
\begin{equation}
J(x,t) = - D \biggl[\frac{\partial P(x,t)}{\partial x} - \frac{c P(x,t)}{D}\biggr]
\label{eq-probcurr} 
\end{equation} 
The expression (\ref{eq-probcurr}) is useful for calculations reported 
in the sections below.

\subsection{Results for the continuum version of the minimal model}

The specific initial condition and the boundary conditions that we 
impose are motivated by the physical situation that our models 
are intended to capture. 
If initial value of $x$ is $x_{0}$, the corresponding initial condition 
for (\ref{main}) can be expressed as 
\begin{equation}
P(x,0)=\delta(x-x_0)
\end{equation}
Throughout this paper we choose $x_{0}=L$, i.e., maximum overlap between 
the outer surface of the MT and innser surface of the coaxial cylindrical 
sleeve-like coupler. The lifetime of an arbitrary kt-MT attachment 
is defined here as the time $t(L)$ taken by the hypothetical particle, 
that is initially at $x=L$, to reach $x=0$ {\it for the first time}. 
Thus, $t(L)$ is essentially a first passage time \cite{redner} that 
fluctuates from one attachment to another. The average value 
$\langle t(L)\rangle$ is the mean life time of the attachment.

Although $x_{0}=L$ initially, $x$ does not necessarily decrease 
monotonically. At any arbitrary instant of time $x$ can also increase, 
just as it can decrease, following the dynamical equations provided 
$0<x<L$. One advantage of the unique initial condition $x_{0}=L$ is 
that the random variation of the lifetime of the kt-MT attachment 
from one run to another arises strictly from the intrinsically random 
kinetics and not from the choice of any random initial condition. 
The distribution of the lifetimes of kt-MT coupler that we have derived 
does not require any further averaging. In contrast, if a random starting
location were selected, one extra averaging (over sufficiently large 
number of random initial locations) would be required to get a meaningful 
distribution of the lifetimes.

Since the MT is not allowed to penetrate the kinetochore plate, and 
the length of the sleeve-like coupler is $L$, the overlap $x$ cannot 
be large than $L$. Therefore, at $x=L$ we impose a reflecting boundary 
condition 
\begin{equation}
J(x,t)|_{(x=L)}=0
\end{equation}
which implies
\begin{equation}
\left.\biggl (\frac{\partial P(x,t)}{\partial x}-\frac{c}{D}P(x,t)\biggr)\right|_{(x=L)}=0
\end{equation}
An absorbing boundary condition 
\begin{equation}
 P(0,t)=0  ~{\rm at}~ x=0
\end{equation}
is imposed so that spontaneous re-formation of the kt-MT is not 
possible and the time taken for $x$ to attain this boundary, 
starting from its initial value $x_0=L$ is the first-passage time 
that we define as the life time of the attachment. 

\begin{figure}[tb]
\begin{center}
\centerline{(a)}
\includegraphics[width=0.65\columnwidth]{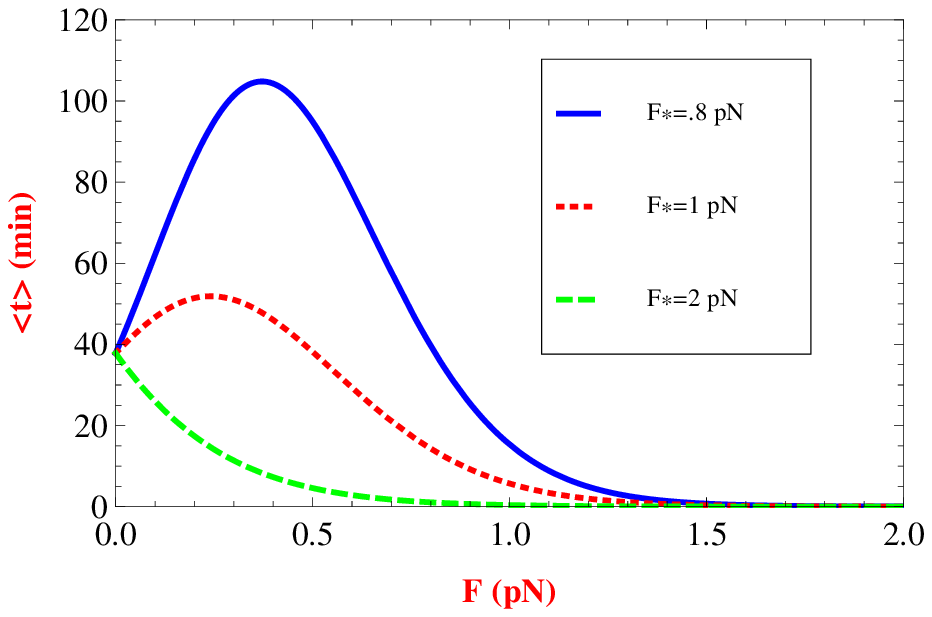}
\centerline{(b)}
\includegraphics[width=0.65\columnwidth]{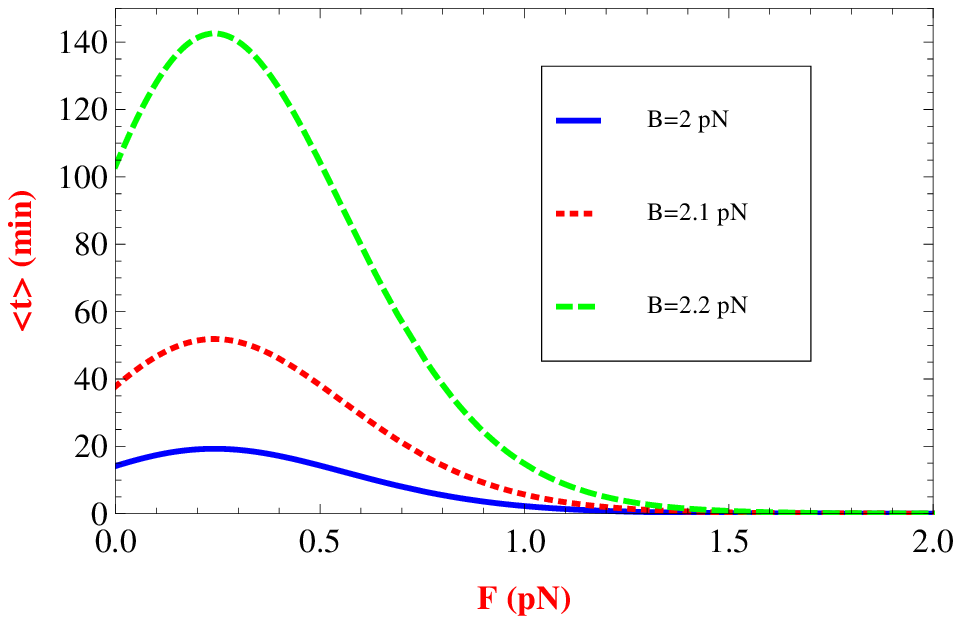}
\end{center}
\caption{Mean life time of the kt-MT attachment is plotted against
the external force $F$, for a few different values of $F_{*}$, (in (a)), 
and $B$ (in (b)) . The parameter values used for 
this plot are $\alpha=30$s$^{-1}$, $\beta_{\text{max}}=350$s$^{-1}$.
}
\label{mean}
\end{figure}
We use the Laplace inversion method \cite{balakrishanan} for extracting 
the relevant quantitative information from equation (\ref{main}).
Defining the Laplace transform of probability density by, 
\begin{equation}
Q(x,s)=\int_{0}^{\infty} P(x,t)e^{-st}dt
\end{equation}
equation (\ref{main}) can be re-expressed in terms of $Q(x,s)$; solving 
that equation under the given boundary conditions (see appendix A for 
the details) we get the exact analytical expression for $Q(x,s)$. 

Let $q(x,s|x_{0})$ denote the probability density, in the Laplace space 
$s$, for the first passage times to reach the position $x$, given that 
the initial position was $x_{0}$. Using the relation \cite{balakrishanan}
\begin{equation}
q(0,s|x_{0})=\frac{Q(x_{1},s|x_{0})}{Q(x_{1},s|0)}
\end{equation} 
between $Q(x,s)$ and $q(x,s|x_{0})$, and the initial condition $x_{0}=L$,   
we get
\begin{equation}
q(0,s|L)=\frac{V\ \exp\biggl(-\frac{cL}{2D}\biggr)}{\biggl[V\ \cosh\biggl(\frac{VL}{2D}\biggr)-c\ \sinh\biggl(\frac{VL}{2D}\biggr)\biggr]}
\label{main2}
\end{equation}
where 
\begin{equation}
V=\sqrt{c^2+u^2}. 
\label{eq-defnV}
\end{equation}
includes contributions from both the drift velocity $c$ and the 
velocity-like quantity 
\begin{equation}
u=\sqrt{4sD} 
\label{eq-defnu} 
\end{equation}
that arises from diffusion.

For the calculation of the mean we do not need the 
distribution in $t$-space.  Instead these can be derived from the 
relation
\begin{equation}
 \langle t \rangle =-\left.\frac{dq(0,s|L)}{ds}\right |_{s=0}
\end{equation}
Using the expression (\ref{main2}) for $q(0,s|L)$, we get 
\begin{equation}
\langle t \rangle =\frac{D}{c^2}\biggl(e^{cL/D}-1\biggr)-\frac{L}{c} 
\label{eq-meant}
\end{equation}

An interesting feature of eqn.(\ref{eq-meant}) is that $cL$ is like an 
effective energy barrier; kt-MT detachment is achieved by crossing this 
barrier. Since in addition to the energetic contributions from $B$ and 
$F$, the barrier $cL$ gets contribution also from $\alpha$ and $\beta$, 
this effective barrier is, at least partly, of kinetic origin. 

External tension $F$ influences the kt-MT attachment time $\langle t \rangle$ 
in two possible ways.\\
(1) $F$ decreases the depth of the linear potential well which 
effectively reduces the mean attachment time.
(2) The same $F$ also reduces the depolymerization rate $\beta(F)$ 
thereby increasing the mean attachment time.\\
These two effects of the same load tension $F$ act against each 
other. The second effect can dominate over the first at small $F$, 
thereby giving rise to nonmonotonic variation of $\langle t \rangle$ with $F$, 
provided $\beta(F)$ falls sharply with increasing $F$; such a 
possibility is ensured if $F_{*}$ is sufficiently small. Therefore, 
as shown graphically in fig.\ref{mean}(a), $\langle t \rangle$ varies 
nonmonotonically with $F$ for both $F_{*}=0.8$pN and $F_{*}=1.0$pN. But, for 
$F_{*}=2.0$pN, the fall of $\beta(F)$ is not sharp enough to 
compensate the reduction of the mean attachment time caused directly 
by $F$. Thus, the physical origin of the nonmonotonic variation of 
$\langle t \rangle$ with $F$ is explained by fig.\ref{mean}(a) in a 
transparent manner. The nonmonotonic variation of $\langle t \rangle$ 
with $F$ is displayed also for all the three different values of the 
parameter $B$ in fig.\ref{mean}(b); 
for any given $F$, the larger is the $B$ (i.e., the deeper is the 
potential well), the longer is the mean life time of the kt-MT 
attachment. 

\begin{figure}[tb]
\begin{center}
\includegraphics[width=.65\columnwidth]{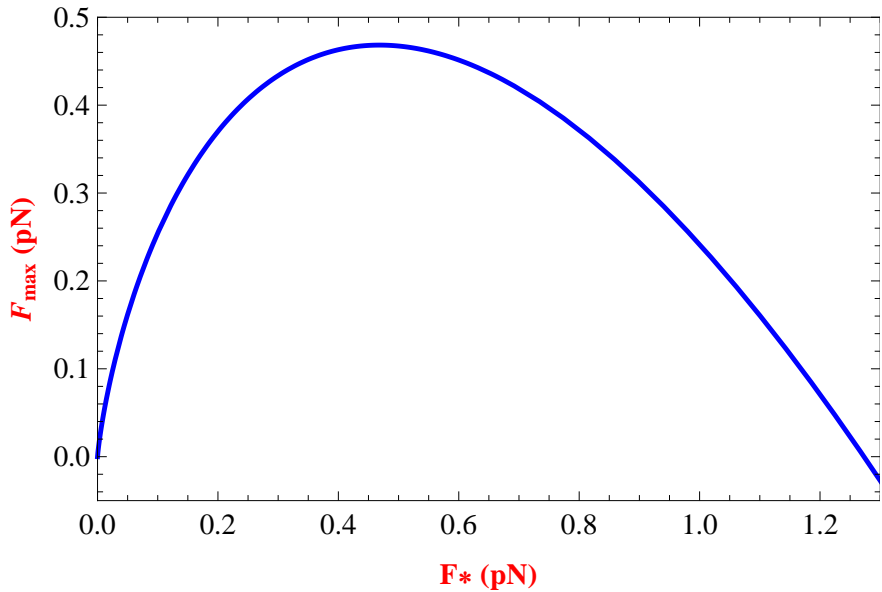}
\end{center}
\caption{Variation of $F_{max}$, the tension that corresponds to the 
maximum if $\langle t \rangle$, is plotted against $F_{*}$. 
}
\label{fig-Fmax}
\end{figure}

Using the symbol $F_{max}$ to denote the tension that corresponds to 
the maximum of the average lifetime, from (\ref{eq-meant}) we get   
\begin{equation}
F_{max}=F_{*}\log_{e}\biggl(\dfrac{k_{B}T\beta \ell}{DF*}\biggr) 
\label{eq-fmax}
\end{equation}
The dependence of $F_{max}$ on $F_{*}$, obtained from (\ref{eq-fmax}), 
is shown in fig.\ref{fig-Fmax}. Interestingly, the variation of 
$F_{max}$ with $F_{*}$ is non-monotonic. 
In the limit $F_{*} \to 0$, $\beta(F) \to 0$ for all $F \neq 0$. 
On the other hand, as $F_{*} \to \infty$, $\beta \to \beta_{max}$ 
for all finite $F$. Therefore, in both these limits, $F$-dependence 
of $\beta$ drops out and kt-MT attachment behaves like a slip-bond. 
The monotonic decrease of 
$F_{max}$ with $F_{*}$ in fig.\ref{mean} arises from the fact that 
we have not plotted $\langle t \rangle$ for even smaller values of 
$F_{*}$ where $F_{*}$ increases with $F_{*}$ because the corresponding 
values of $\langle t \rangle$ turned out to be unphysically long. 

At this stage we can scrutinize the necessity of the exponential form 
(\ref{eq-Fdepbeta}) for the tension-dependence of the MT depolymerization 
rate $\beta$. As shown in fig.\ref{mean}(a), a catch-bond-like behavior 
follows if $\beta(F)$ decreases sufficiently sharply with increasing $F$.
On the other hand, if $\beta(F)$ falls very slowly with increasing $F$ 
the system exhibits a slip-bond-like behavior. The exponential decrease 
assumed in equation (\ref{eq-Fdepbeta}) is an ideal choice for the 
$F$-dependence of $\beta(F)$ because the crossover from catch-bond to 
slip-bond can be displayed by simply varying $F_{*}$ which determines 
the sharpness of the decrease $\beta(F)$.

The series representation of equation (\ref{eq-meant}) 
\begin{equation}
\langle t \rangle = \biggl(\frac{L^2}{2D}\biggr)\biggl[1+2\sum_{n+1}^{\infty} \frac{1}{(n+2)!}\biggl(\frac{cL}{D}\biggr)^{n}\biggr]
\label{eq-approxmeant}
\end{equation} 
is exact. In the special case $B-F=0, \alpha-\beta=0$ 
eqn.(\ref{eq-approxmeant}) reduces to $\langle t \rangle = L^2/(2D)$ which, 
as expected, arises from kt-MT detachment caused by pure diffusion. 
The corrections to the diffusive result in the limit $cL/D << 1$, can 
be estimated from eqn.(\ref{eq-approxmeant}). 

As we clearly stated in the introduction, the aim of the minimal model 
(and, to some extent, even the more detailed model) is not quantitative 
comparison with the experimental data reported by Akiyoshi et al. Instead, 
our aim is to present analytical calculations for a minimal model that 
provides physical insight into the interplay of opposing forces and 
competing kinetic processes that, together, determine the stability of 
the kt-MT coupler. Nevertheless, for graphical plots, which often provide 
a more intuitive understanding, one needs numerical values of the parameters.
The largest strength of the load tension applied by Akiyoshi et al. 
\cite{akiyoshi10} was about $13$ pN while the lifetime peaked closed to  
$4$pN. The longest life time that they measured was about $50$ minutes. 

Although the mean lifetimes plotted in fig.\ref{mean}(a) are comparable 
to the measured values for at least one set of parameter values (corresponding 
to $F_{*}=1$pN), the numerical values of the corresponding load tension $F$ 
are much smaller than those applied in the experiment of Akiyoshi et al. 
\cite{akiyoshi10}. However, this discrepancy does not invalidate our theory. 
The magnitudes of the forces in our model depend on the numerical values 
of the parameters like, for example, $B$, $\beta(0)$, etc. On the other 
hand, the corresponding effective values in the experiments of Akiyoshi 
et al.\cite{akiyoshi10} are not known. Therefore, in our plots we have 
used parameter values that yield lifetimes comparable to those reported 
in experiments. Most probably, our minimal model does not capture all the 
details of the in-vitro study \cite{akiyoshi10}. Even so, we note that the 
inclusion of our two postulates is sufficient to generate non-monotonic 
variation of the lifetimes with load tension.

\section{The minimal model: discrete formulation and results}
\label{sec-minimaldisc}

We begin by pointing out that in the continuum formulation treated above, 
the expression (\ref{eq-probcurr}) for the probability current density 
$J(x,t)$ can be recast as 
\begin{equation}
J(x,t) = - D \biggl(\frac{\tilde{U}'}{k_BT} P + \frac{\partial P}{\partial x}\biggr)
\end{equation}
with $\tilde{U}'(x) = d\tilde{U}/dx$ where the effective potential 
$\tilde{U}$ is given by 
\begin{equation}
\frac{\tilde{U}(x)}{k_BT}=\biggl[\frac{(F-B)}{k_{B}T}+\ell\frac{\beta-\alpha}{D}\biggr] x.
\end{equation}
Note that $\tilde{U}(x)$ accounts for the drift caused by the force $B-F$ 
as well as that arising from the polymerization-depolymerization kinetics 
of the MT. 

In this section, we utilize the effective potential $\tilde{U}(x)$ to 
simulate the stochastic movements of kinetochore by using a method of 
discretization popularized by Wang, Peskin and Elston (from now onwards 
referred to as WPE method) \cite{wang03,wang07}. 

\subsection{Discrete formulation of the minimal model}

\begin{figure}[tb]
\begin{center}
\includegraphics[width=.9\columnwidth]{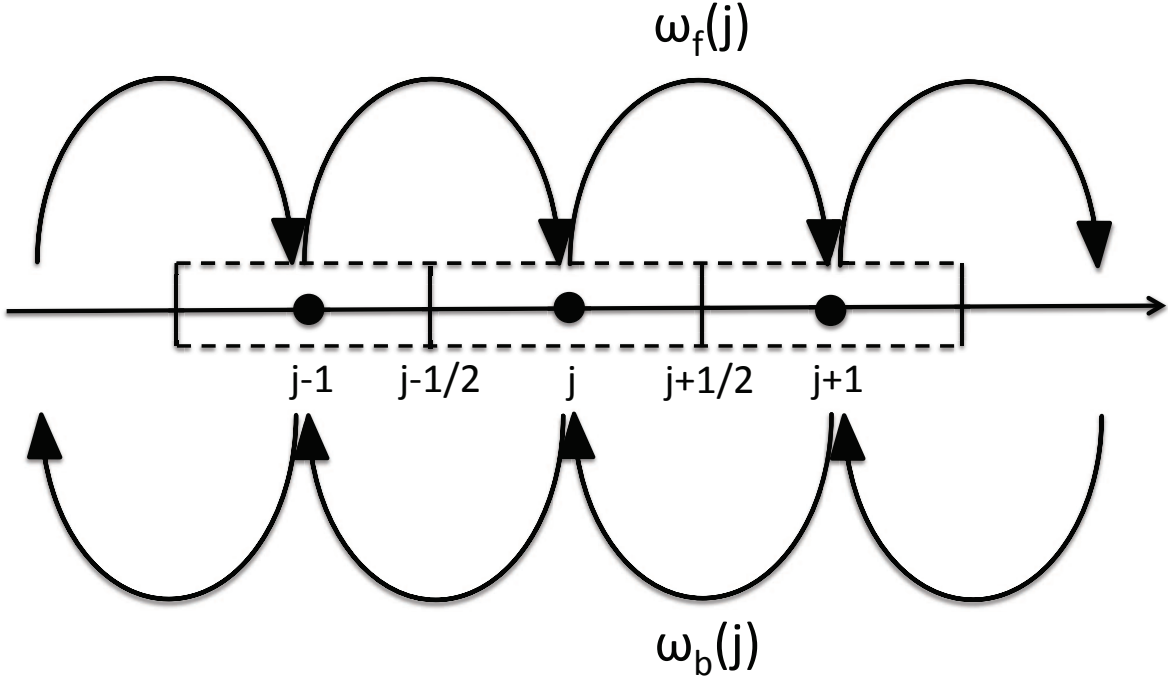}
\end{center}
\caption{The continuous one-dimensional space of length $L$ is discretized 
into $M$ cells each of length $h=L/M$ following the WPE prescription (see 
text for details).
}
\label{fig-discretize}
\end{figure}

Following WPE prescription, we discretized the space into $M$ cells, 
each of length $h = L/M$; however, $h$ need not be identical to $\ell$, 
the separation between the two consecutive binding site of a MT 
(see fig.\ref{fig-discretize}). Accordingly, the continuous effective 
potential $\tilde{U}(x)$ is replaced by its discrete counterpart 
\begin{equation}
\frac{\tilde{U}_{j}}{k_BT}=\biggl[\frac{(F-B)}{k_{B}T}+\ell\frac{\beta-\alpha}{D}\biggr] x_{j}
\label{eq-discreteU}
\end{equation}
where $x_j$ denotes the position of the center of the $j$-th cell. 

Next we approximate the continuous movement described by the Fokker-Planck 
equation (\ref{main}) by a master equation in terms of discrete jumps 
from the center of one cell to that of an adjacent cell either in the 
forward or backward direction. In this case the expressions for the 
forward and backward transition rates $\omega_{f}(j)$ and $\omega_{b}(j)$ 
(see fig.\ref{fig-discretize}) are given by \cite{wang03} 
\begin{eqnarray}
\omega_{f}(j)&=&\frac{D}{h^2}\frac{-\frac{\delta \tilde{U}_{j}}{k_{B}T}}{\exp\biggl(-\frac{\delta \tilde{U}_{j}}{k_{B}T}\biggr)-1} 
= \dfrac{1}{h}\dfrac{\dfrac{B-F}{\Gamma}+\ell(\alpha-\beta)}{\exp\biggl(-\dfrac{\delta \tilde{U_{j}}}{k_{B}T}\biggr)-1}
\label{wf}
\end{eqnarray} 
\begin{eqnarray}
\omega_{b}(j)&=&\frac{D}{h^2}\frac{\frac{\delta \tilde{U}_{j}}{k_{B}T}}{\exp\biggl(\frac{\delta \tilde{U}_{j}}{k_{B}T}\biggr)-1} 
= \dfrac{1}{h}\dfrac{\dfrac{F-B}{\Gamma}+\ell(\beta-\alpha)}{\exp\biggl(\dfrac{\delta \tilde{U_{j}}}{k_{B}T}\biggr)-1}
\label{wb}
\end{eqnarray}
where 
\begin{equation}
\delta \tilde{U}_{j}= \tilde{U}_{j+1}-\tilde{U}_{j}
\end{equation} 
For the simple potential (\ref{eq-discreteU}), $\omega_{f}(j)$ and 
$\omega_{b}(j)$ are independent of site index $j$.\\

\begin{figure}[tb]
\begin{center}
\includegraphics[angle=-90,width=0.9\columnwidth]{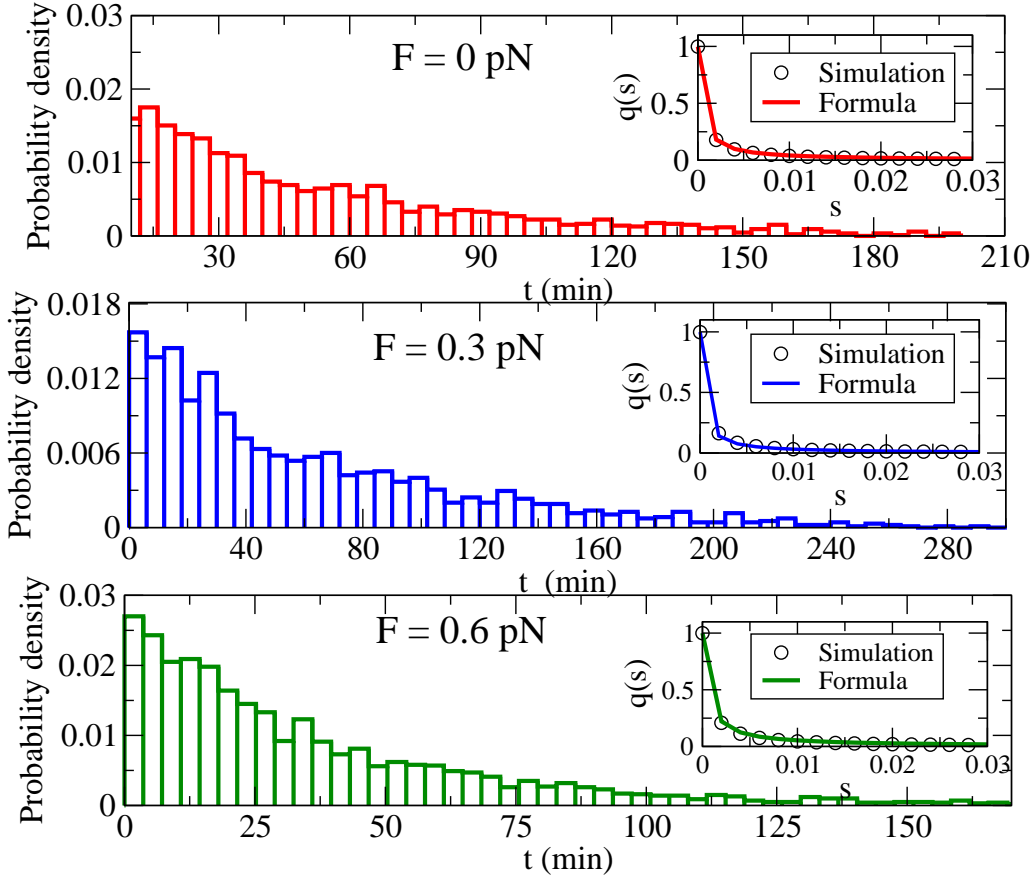}
\end{center}
\caption{Probability density (histogram) of the distribution of 
lifetimes of the kt-MT attachments obtained from direct simulation 
of the discrete version of our minimal model is plotted for a few 
different values of external tension $F$. The same Laplace-transformed 
distributions are plotted in the corresponding insets by discrete 
symbols ($\circ$). Our theoretically predicted distributions in 
Laplace space for each of the three values of $F$, obtained from 
the analytical expression (\ref{main2}),  are also shown by the 
continuous lines in the corresponding insets. 
The numerical values used in this figure for the parameters are:
$\alpha=30$s$^{-1}$, $\beta_{\text{max}}=350$s$^{-1}$, $B=2.1$pN, $F_{*}=1$pN, $L=50$nm 
$D=700$nm$^{2}$/s$^{-1}$.}
\label{hist}
\end{figure}

\subsection{Results for the discrete version of the minimal model}

In computer simulations, the MT is fully inserted into the sleeve in 
the initial state. The simulation produce the numerical data for the 
first passage time of the MT tip to exit the coupler. In fig. 
\ref{hist}, histograms of these numerical data are plotted for a few 
values of force $F$. Laplace transform of these distributions 
are compared in the insets with the corresponding prediction of the 
exact analytical formula (\ref{main2}). Excellent agreement between 
the theoretical prediction and the simulation data (albeit in the 
Laplace space) demonstrates how the theory may be useful in analyzing 
also the distribution of experimentally measured lifetimes of kt-MT 
attachments {\it in-vitro}.

\section{Beyond the minimal model: effects of friction in the absence of motor proteins} 
\label{sec-realistic}

\begin{figure}[tb]
\begin{center}
\includegraphics[width=0.55\columnwidth]{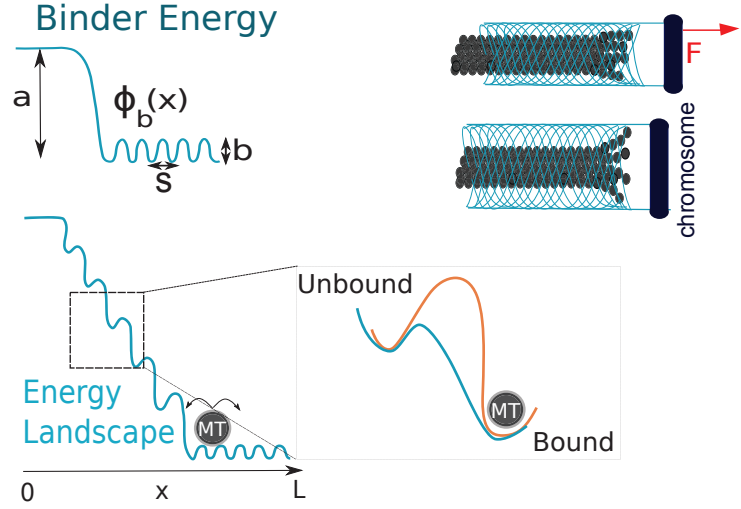}
\end{center}
\caption{ The theoretical model, extended by incorporating a realistic kt-MT interaction potential, is depicted schematically (see
text for the details of $\psi_{b}(x)$ and $\Psi_{b}(x)$). The load tension $F$ tilts the potential energy landscape $\Psi_{b}(x)$. The tension-dependent depolymerization rate $\beta(F)$ increases the effective barrier along the unbinding pathway (orange line, inset) by slowing down depolymerization.}
\label{fig2a}
\end{figure}

In this section we begin with the assumption that the inner surface of 
the cylindrically shaped coupler consists of only passive binders. 
We assume that each binder head engages with the MT by obeying a unit 
energy function $\phi_{b}(x)$ (see fig.\ref{fig2a}), which has two key 
parameters: 
$a$ measures free energy  drop due to binder affinity for the MT 
lattice, and $b$ describes the activation barrier for transitions between 
specific MT lattice binding sites \cite{shtylla11}. In other words, 
$a$ is a measure of the strength of the MT-binder affinity while $b$ 
is a measure of the ``roughness'' or friction of the MT-coupler interface. 
The total potential energy function is given by (see fig.\ref{fig2a})
\begin{equation}
\Psi_{b}(x)=\sum_{n=0}^{N_{b}-1}\phi_{b}(x-ns)
\end{equation}
where $s$ is the spacing between consecutive coupler binders 
(see Fig \ref{fig2a}). Binder spacing is an arbitrary parameter. 
Here we set $s=\ell$, where $\ell$ is the distance between consecutive 
MT binding sites.

\begin{figure}[tb]
\begin{center}
\includegraphics[width=0.55\columnwidth]{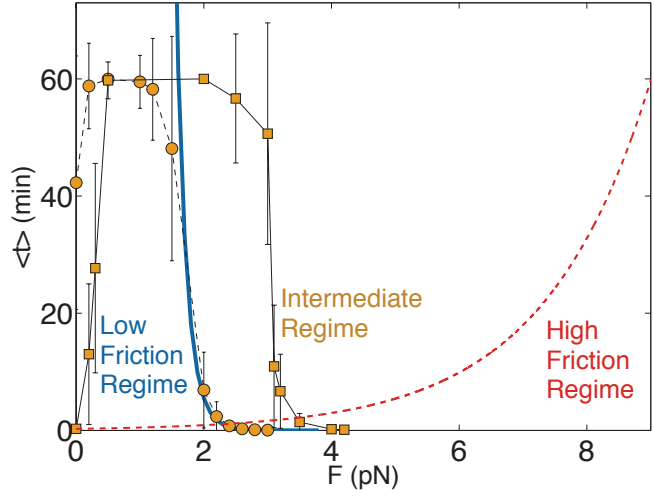}
\end{center}
\caption{ Mean attachment time for the model depicted in fig.7 plotted for three different regimes of parameters. Low-friction
regime (blue line) is obtained from eq. (20) with a = 0.5k$_B$T, b = 0.001a. Intermediate regime (circle) a = 0.5k$_B$T, b = 0.04a,
and intermediate regime (square) a = 0.6k$_B$T, b = 0.2a. High-friction regime (dashed red line) is obtained from eq. (32) with
$\beta_{max} = 300$, $\alpha = 0$, $F^{*} = 0.6$ pN.}
\label{fig2b}
\end{figure} 

For $\Psi_{b}(x)$ in our numerical calculations, we have used the simple 
expression 
\begin{align}
\Psi_{b}(x)&=
\begin{cases}
f(x)\left(1-\cos(\frac{2\pi x}{\ell})\right)+h(x)&x\leq N_{b}{\ell}\\ \label{eq5}
f(N_{1}\ell)\left(1-\cos(\frac{2 \pi x}{\ell})\right)+h(N_{b}\ell)&x>N_{b}{\ell}.
\end{cases}
\end{align}
where $f(x)=\dfrac{a}{2\ell}\left(\frac{b}{a}x+C\right)$, 
$C=0.172$ and $h(x)=-\dfrac{ax}{\ell}$. The linear and scalar 
coefficients in eq.~(\ref{eq5}) arise because we use a Fourier 
series to approximate the well function expression. Thus, the 
MT-coupler interaction is represented more realistically 
by $\Psi_{b}(x)$ than by $U_{b}(x)$ of the minimal model. 
The friction arising from the potential landscape $\Psi_{b}(x)$ 
gives rise to a jump-drift-diffusion process.  As we'll point 
out below, this general version of our model reduces to the 
minimal version discussed above, and hence effectively to a 
drift-diffusion process, in a special limiting case.

In infinitesimal time interval $dt$, the change $dx$ of the overlap is 
given by
\begin{equation}
dx(t)= \frac{1}{\Gamma}\left[ -\Psi_{b}'(x)-F\right]dt+ \ell dN_{r}(t)+\sigma dW(t) 
\label{eq1}
\end{equation}
where, as before, the constant $F$ is the external load tension and 
$\Gamma$ is the effective drag coefficient. $dW(t)$ accounts for the 
thermal diffusion of the coupler on the lattice with 
$\sigma=\sqrt{2k_{B}T/\Gamma}$, and $N_{r}(t)$ is a Poisson counting 
process describing the MT tip dynamics with rate $r=\alpha-\beta$ that 
is decided by the difference of the MT polymerization rate $\alpha$ 
and MT depolymerization rate $\beta$. 
The stochastic differential equation (\ref{eq1}) is essentially equivalent 
to the Langevin equation (\ref{eq-langevin}); the form (\ref{eq1})  
is used in our numerical computations for the convenience of implementation.

$N_{b}$ is characteristic of the structure of the coupler (coupler length) whereas its energetics depend on $\Psi_{b}$ (i.e., on the parameters $a$, $b$) and $F$; the stochastic kinetics are influenced by the interplay of forces arising from the potential landscape, random Brownian forces, and by $\alpha$, $\beta$. By a combination of standard analytical and numerical methods, we 
study the trends of variation of $\langle t(L) \rangle$ with (a) $F$, as 
well as, the (b) size, (c) energetics, and (d) kinetics of the coupler.

It is difficult to derive an exact analytical expression for the mean first passage time corresponding to eq.~(\ref{eq1}) that would be valid in all parameter regimes. Therefore, we explore two limiting cases for which explicit approximate solutions can be obtained: (a) {\it Low-friction regime} ($b<<k_{B}T$), and (b) {\it High-friction regime} ($b>>k_{B}T$). 
In the {\it low-friction regime} the coupler can easily rearrange its position relative to the MT; in this regime $\langle t(L) \rangle$ approximated well by the expression in equation (\ref{eq-meant})
In the {\it high-friction regime} large $b$ leads to strong local pinning that practically stalls it. The only mode available for detachment of the coupler is via depolymerization of the MT. Consequently, in this regime
\begin{equation}
\langle t(L) \rangle \approx \biggl(\frac{L}{\ell\beta_{\text{max}}}\biggr)~ e^{F/F_{*}}.
\label{eq-bindTL}
\end{equation}
The exponential increase of $\langle t(L) \rangle$ with $F$ in the 
high-friction regime can be regarded as an artifact of the unphysical 
(or, unrealizable) nature of this regime.

Unless explicitly stated otherwise, the parameter values that we have used 
for our numerical calculations in this section are given in the table 
\ref{default}. 

\begin{table}[h!]
\begin{center}
\begin{tabular}{ccc}
Parameter Description & Symbol & Values Tested \\
\hline
\hline \cr
MT binding site spacing & $\ell$ & 8/13 nm \cite{hill85,joglekar02,shtylla11}\cr
Maximal coupler length &$L$& 50 nm \cite{joglekar06,Gonen2012,Joglekar2008,Johnston2010}\cr
Maximal number of coupler binders &$N_{b}$& 15-65 \cite{Powers2009,Johnston2010}\cr
Polymerization Rate & $\alpha$ & $20-50$ $s^{-1}$\cite{joglekar02,hill85,Waters1996,shtylla11}\cr
Maximal Depolymerization Rate & $\beta_{\text{max}}$ & $100- 350$ $s^{-1}$ \cite{hill85,Waters1996,shtylla11}\cr
Critical Depolymerization Force   & $F_{*}$ & $0.3-5$ pN (estimated here)\cr
External load Force   & $F$ & $0-5$ pN \cr
Force of attractive kt-MT interaction  & $B$ & $2-2.5$ pN \cr
MT lattice/binder binding energy & $a$ & $0.4k_{B}T-3 k_{B}T$ \cite{hill85,joglekar02,Powers2009,shtylla11}\cr
Binder activation barrier   & $b$ & $0.001a-0.4a$ (estimated here)\cr
Diffusion constant  & $D$ & $700$ nm$^2$s$^{-1}$ \cite{hill85,joglekar02,shtylla11}\cr
Effective drag coefficient & $\Gamma$ &  $6$ pNs$\mu$m$^{-1}$\cite{hill85,Marshall2001,joglekar02,shtylla11}\cr
\hline
\end{tabular}
\end{center}
\label{default}
\end{table}%

\begin{figure}[tb]
\begin{center}
\includegraphics[width=0.6\columnwidth]{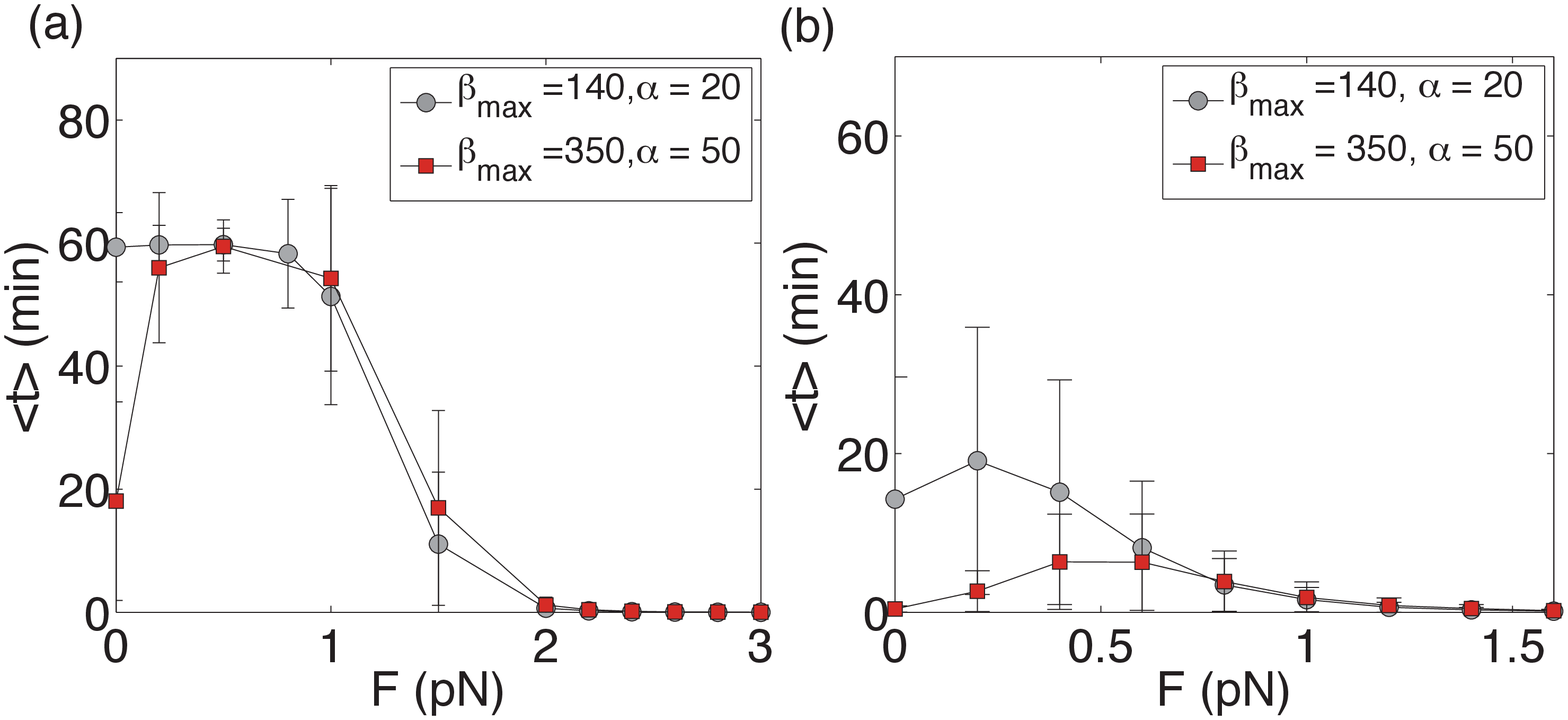}\\
\includegraphics[width=0.6\columnwidth]{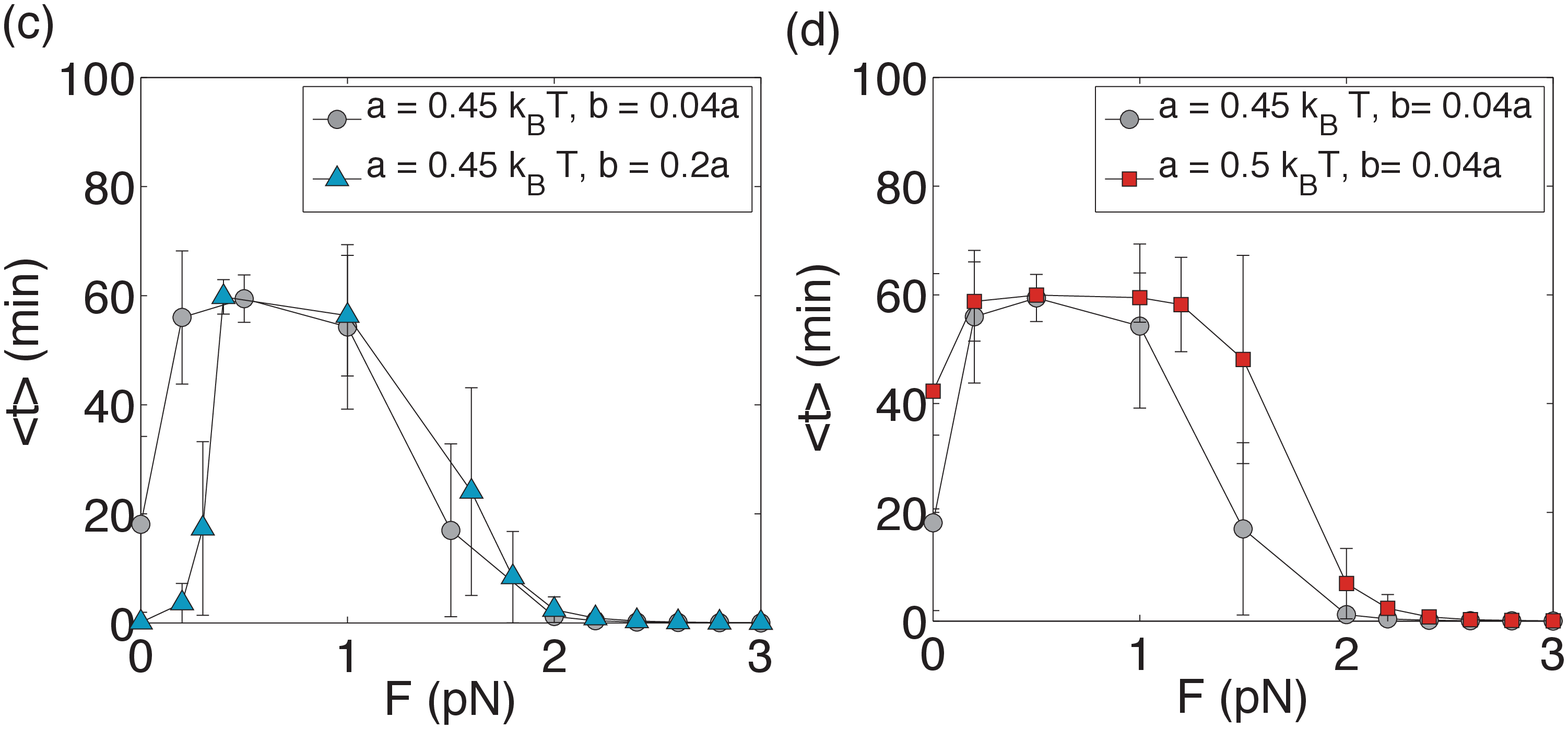}\\
\includegraphics[width=0.45\columnwidth]{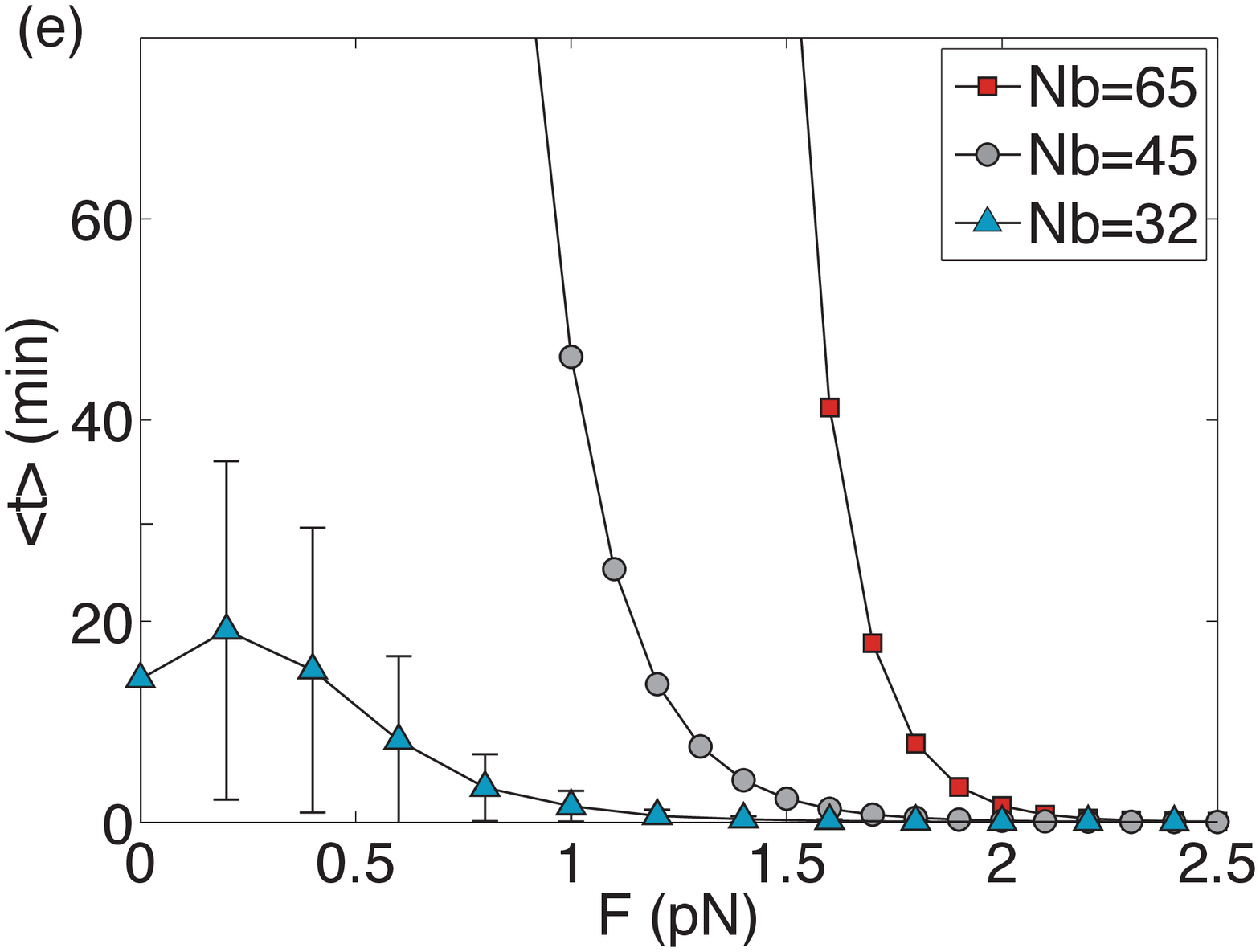}\\
\end{center}
\caption{ Results of simulations for the model depicted in fig.7, using equation (31) and $F^{*}= 1/3$. (a)-(b) $\langle t(L) \rangle$ for two sets of maximal rates of MT depolymerization with the ratio $\beta_{max}/\alpha$ fixed. In (a) the data are obtained for an {\it intermediate} regime coupler, with $a = 0.45$ k$_B$T, $b = 0.04a, N_{b}=45$, and in (b) for a shorter coupler in the {\it low-friction} regime with $a = 0.4$ k$_B$T , $b = 0.001a, N_{b}= 32$. (c)-(d) $\langle t(L) \rangle$ for {\it  intermediate} regime couplers. In (c) are shown data with two different strengths of friction caused by two values of $b$, and in (d) for two different binding energies given by $a$. The MT rates are  $\beta_{max}= 350$ s$^{-1}$, $\alpha= 50$ s$^{-1}$ for (c)-(d).
(e) $\langle t(L) \rangle$ plotted against $F$ for three different lengths of the {\it low-friction} coupler from (a) with $a = 0.4$ k$_B$T , $b = 0.001a$, $\beta_{max}= 140$ s$^{-1}$, $\alpha= 20$ s$^{-1}$. Error bars mark standard deviation.}
\label{fig3}
\end{figure}

\begin{figure}[tb]
\begin{center}\includegraphics[width=3.65in,height=2.65in]{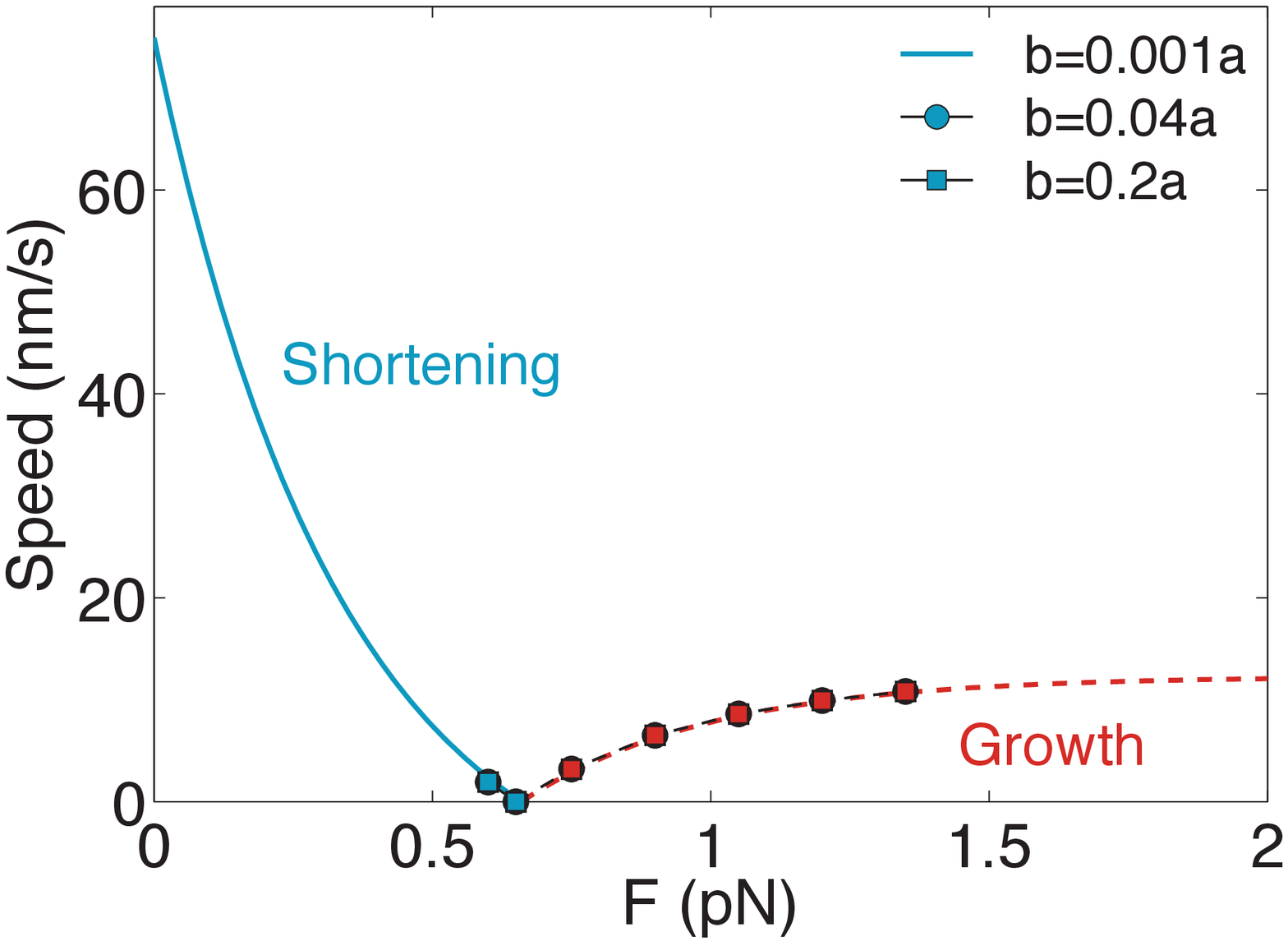}
\caption{Force-speed relations obtained by direct numerical simulation of the model depicted in fig.7, using the corresponding
dynamical equation (31) and $F^{*} = 1/3$, is plotted in the low-friction regime and the corresponding data for the intermediate
regime. The parameters are $N_{b}$ = 65, a = 0.45 k$_B$T, $\beta_{max}= 140$ s$^{-1}$, $\alpha= 20$ s$^{-1}$. Error bars mark standard deviation.}
\label{fig3d}
\end{center}
\end{figure}

For numerical simulations of this version of our model in the intermediate 
regime we used the parameter values which are within the range reported in 
the literature\cite{joglekar09,wan09,lawrimore11}. Consequently, computed 
$\langle t(L)\rangle$ turn out to be comparable to those observed in the 
experiments.  Some of the typical curves are plotted in Figs.\ref{fig3}. 

\noindent $\bullet$ {\it Effects of variation of rate of kinetics}

The effects 
of varying the maximal rate of depolymerization $\beta_{\text{max}}$, 
keeping the ratio $\beta_{\text{max}}/\alpha$ fixed, is demonstrated in 
figs.\ref{fig3}(a) and (b). Clearly, in such situations, the kt-MT 
attachment can exhibit catch-bond-like or slip-bond-like behavior 
depending on the value of $\beta_{\text{max}}$. 

\noindent $\bullet$ {\it Effects of variation of parameters that characterize energetics}

The effects of the energetics, namely the influences of the parameters $a$ 
and $b$ that determine the potential energy landscape, are displayed in 
figs.\ref{fig3}(c) and (d). The effects of varying the number of binders 
$N_{b}$ is shown in fig.\ref{fig3}(e). In all these cases the nonmonotonic 
variation of $\langle t(L) \rangle$ with $F$ seems to be a generic feature, 
except that $\langle t(L) \rangle$ may become too large to be observable 
for some values of the parameters (see, for example, fig.\ref{fig3}(e).

\noindent $\bullet$ {\it Tension-induced switching from depolymerization to polymerization}

Fig.\ref{fig3d}  shows the key role of the external tension $F$ in 
deciding the direction of the velocity. 
Suppose the parameter values are chosen such that $\beta_{\text{max}}$ 
dominates over $\alpha$ in the absence of external tension thereby 
ensuring depolymerization of the MT. When the external tension is applied, 
the depolymerization is suppressed. Increasing $F$ can eventually cause 
such strong suppression of the depolymerization that, beyond a certain 
strength of $F$, $\alpha$ dominates over $\beta(F)$. In such a scenario, 
the MT switches from the depolymerization mode to polymerization mode at a 
threshold strength of $F$.

\section{Summary and Conclusions}

In this paper, we have presented theoretical models of kt-MT attachment 
at three levels of detail starting with a minimal version that captures 
only the key components and essential processes. 
These models mimic kt-MT attachments reconstituted {\it in-vitro} from 
purified components extracted from budding yeast \cite{akiyoshi10}. 

Exploiting the simplicity of the minimal version of the model, we have 
calculated the distribution of the lifetimes of the kt-MT attachments 
analytically and compared the results with the corresponding simulation data. 
Equation (\ref{main2}) encapsulates the main result of this part of the 
manuscript. The strength of the minimal 
model is that the results can be derived analytically. In spite of its 
simplicity, this minimal model provides deep insight into the roles of 
opposing forces and competing kinetics in the tension-dependence of the 
life time of the kt-MT attachments. 

The second version of the model is an extension of the minimal model; 
it incorporates some more realistic features of the interaction between 
the MT and the coupler. For the second model we have succeeded in 
calculating analytically only the mean lifetime of the kt-MT attachments 
under reasonable schemes of approximation.

The analytical expression (\ref{eq-meant}) for the mean life time of the 
kt-MT attachment in the minimal model has been derived from the 
corresponding full distribution (\ref{main2}) of the life times, 
The nonmonotonic variation of the mean lifetime of the kt-MT attachments 
with external tension exhibited over wide range of values of the model 
parameters in both the models is consistent with the similar trend observed 
in the {\it in-vitro} experiments \cite{akiyoshi10}. The realistic potential 
landscape in the second model does not alter this qualitative trend. 
We have also demonstrated that the nonmonotonicity of the variation of 
$\langle t \rangle$ with $F$ depends on the rapidity of decrease of $\beta(F)$  
with increasing $F$. By varying $F_{*}$, which determines the sharpness 
of the decrease of $\beta(F)$, we have shown how the catch-bond-like 
behavior crosses over to a slip-bond-like behavior. Moreover, our 
analytical calculations also show that, unlike other conventional catch-bonds, 
the catch-bond-like behavior of the kt-MT attachment arises from the 
interplay of not only forces derivable from the potential landscape, but also 
by a competition of the kinetics of polymerization and depolymerization of 
the MT. In fact, the barrier against the breakdown of the attachment gets 
contributions from both opposing forces and competing kinetic processes.  
The slip-bond-like monotonic decrease of $\langle t(L) \rangle$ with $F$ 
observed in some of the parameter regimes (whose physical origin has been 
explained in terms of the nature of $F$-dependence of the depolymerization 
rate $\beta(F)$), might be detectable in experiments under conditions 
different from those used by Akiyoshi et al.\cite{akiyoshi10}.

As expected, we have found that the molecular motors enhance stability or 
tend to destabilize the kt-MT attachment depending on whether the force it 
generates opposes or assists the external tension. The latter awaits 
experimental confirmation. Moreover, our analytical predictions on the 
distribution of the lifetimes will be useful in the future in analyzing the 
statistical properties of the lifetimes provided the {\it in-vitro} 
experiment on the reconstituted kt-MT attachment is repeated sufficiently 
large number of times. 

In principle, the models presented here can 
be further extended by including active force generators like molecular 
motors \cite{chowdhury13a,chowdhury13b}. However, so far such active force 
generators have not been detected at the kt-MT interface in budding yeast 
although such force-generating motors have been found in other cells 
including, for example, mammalian cells. But, in contrast to budding 
yeast, where each kinetochore can attach to only one MT, multiple MTs 
can attach with each kinetochore in a mammalian cell. Since modeling of 
kt-MT attachment in this paper is focussed almost exclusively on 
budding yeast, we have not discussed the effects of molecular motors in 
the main text. Thus, the extended model in appendix B is not a complete 
model of kt-MT attachment in mammalian cells; it is intended to indicate 
how the effects of molecular motors can be incorporated within the broad 
theoretical framework developed in this paper.
In near future we hope to extend our model to kt-MT attachments in 
mammalian cells. However, at present, the main hurdle in this modeling 
is the lack of our current understanding as to how multiple MTs attached 
to the same kt coordinate their kinetics \cite{mcintosh12}.

\subsection{Comparison with earlier kinetic model of kt-MT attachments}
\label{sec-oldmodels}

The ``sleeve model'' developed originally by Hill 
\cite{hill85} was adapted by Joglekar and Hunt \cite{joglekar02}, to 
account for the phenomenon of ``directional instability'' 
\cite{skibbens93,cassimeris94} (also 
referred to as chromosome oscillation) observed  
during mitosis in vertebrates. In those cells each kt can attach with 
several MT (typically, up to 35) and this feature was also incorporated 
in the model studied by Joglekar and Hunt by carrying out computer 
simulations.

Our work differs significantly from that of Joglekar and 
Hunt \cite{joglekar02}. First, our model is intended to account for the 
qualititative features of the results obtained from a much simpler 
{\it in-vitro} system where only one MT is attached to only a single kt 
that is subjected to an externally applied load force that, in our model, 
is assumed to alter the rate of MT depolymerization itself. Moreover, 
because of their motivation in studying directional instability, Joglekar 
and Hunt \cite{joglekar02} put major emphasis on monitoring the 
{\it spatial} displacements that characterize this instability. In 
contrast, the quantity of our main interest is the life {\it time} of 
the kt-MT attachments. 

The spatial excursions of the kinetochores was also the main quantity 
of interest of Gardner et al.\cite{gardner05} who studied the effects 
of tension-mediated regulation of the kinetics of kinetochore MTs by 
computer simulation of a related model. The scope of the more detailed 
model studied by Scholey et al.\cite{civelekoglu06} was even broader. 
Our force-balance equations are similar, at least in principle, to the 
force-balance equations formulated by Scholey et al.\cite{civelekoglu06}. 
However, instead of the spatial displacements of the kinetochores, the 
lifetimes of the kt-MT attachments are the main focus of our investigation. 

Most of the earlier models were based on the assumption that the coupler 
is a stable sleeve or ring. In recent years, a new class of models have 
been developed on the basis of an altogether different assumption regarding 
the nature of the coupler. In this scenario \cite{zaytsev13} 
microtubule associated proteins (MAPs) are assumed to make transient 
attachments with the MT (i.e., each MAP attaches to the MT for a brief 
duration before getting detached from it). The entropic contribution to 
the force is significant and slight variation in the mean lifetime of the 
individual transient attachments of the MAPs can have large effect on 
the lifetime of the kt-MT attachment.
  
In order to account for their experimental observations, Akiyoshi et al. 
\cite{akiyoshi10} developed a 2-state kinetic model. This model 
was motivated by the formal analogy with the models of `catch-bonds' in 
other systems. The growing and shrinking states of the MT were argued 
to be analogs of the strongly- and weakly-bound states in the models of 
catch-bonds. In their 4-parameter kinetic model, two parameters were the 
rates of transitions between the states of growth and shrinkage of the 
MT. The remaining two parameters were the rates of detachment of the 
kinetochore from the MT while the latter is in the growing and shrinking 
phases, respectively. Postulating exponential dependence of the type 
$k_n(F)=k_n^{0} \exp(F/F_n)$ for each of the four rate constants (n=1,2,3,4), 
on the external tension $F$, Akiyoshi et al. extracted the numerical 
values of the 8 parameters $k_n^{0},F_n$ (n=1,2,3,4) corresponding to 
the best fit to their experimental data \cite{akiyoshi10}. 

In their model, Akiyoshi et al.\cite{akiyoshi10} did not explicitly treat 
the kinetics of the polymerization and depolymerization, by attachment 
and detachment of the successive subunits (more precisely, the 
$\alpha-\beta$ tubulin dimers), during the growing and shrinking phases 
of the MT, respectively. Therefore, the force-dependence of the four 
rate constants that provide the best fit to their experimental data do 
not indicate the corresponding force-dependences of the rates of 
polymerization and depolymerization of the MT. Moreover, the potential 
energy of interaction between the MT and the coupler as well as the effect 
of the external tension on this potential energy landscape are not 
incorporated explicitly in their model. Thus, the model of Akiyoshi et 
al.\cite{akiyoshi10} does not directly demonstrate the interplay of the 
opposing forces and competing kinetics. 

Since neither any structural features of the coupler, nor the nature of 
its interaction with the MT enters explicitly into Akiyoshi et al.'s 
\cite{akiyoshi10} model, their model cannot be used to study the effects 
of the size and composition of the coupler or that of the nature of the 
MT-coupler interaction. Neither can it be used to account for the special 
features of hybrid couplers. In contrast, in this paper we have used our 
models to study the effects of varying (i) the size of the coupler, (ii) 
relative population of active force generators (molecular motors) and 
passive binders, as well as (iii) the depth and roughness of the potential 
of interaction between the coupler and the MT.

\subsection{Comparison with catch-bond mechanisms in other ligand-receptor systems}
\label{sec-ligrecep}

Catch bond formed by the fimbriae of E-coli bacteria with target cell 
surface  and that formed by eukaryotic P-selectin and integrin receptors 
with their ligands have been studied extensively in recent years 
\cite{thomas08a,thomas08b,thomas09,sokurenko08,prezhdo09,evans07,zhu05,zhu08,mcever10,sundd11,sundd13} 
to understand their mechanism. 

In this paper we have focussed exclusively on the kt-MT attachments to 
understand the mechanism of its catch-bond-like behavior \cite{akiyoshi10} 
using simple theoretical models. One distinct feature of the MT is that, 
unlike other ligands, these stiff filaments are capable of generating push 
and pull by their polymerization and depolymerization, respectively. 
In particular, the pulling force generated by a MT is powered by the 
hydrolysis of GTP bound to the tubulin subunits. As we have shown in 
this paper, the competing polymerization and depolymerization of MT 
gives rise to a unique feature of the catch-bond-like behavior of 
kt-MT attachments where the stable state can be a non-equilibrium (local) 
steady state rather than a local minimum of the free energy of the 
ligand-receptor system. We hope the modeling methodology developed here 
can be usefully adapted to understand the catch-bond-like behavior of 
depolymerizing actin under load tension observed recently by 
carrying out {\it in-vitro} experiments with atomic force microscope (AFM) 
\cite{lee13}.

\noindent {\bf Acknowledgements:}  DC thanks Frank J\"ulicher for valuable 
comments. This work has been supported in part by CSIR (India) through a 
senior research fellowship (AKS), by NSF through the grant DMS- 1358932 (BS), 
by MBI, the Ohio State University, through the NSF grant DMS 0931642 (BS and 
DC), by IIT Kanpur through Dr. Jag Mohan Garg Chair professorship (DC), by 
SERB (India) through J.C. Bose National Fellowship (DC), by DBT (India) 
through a research grant (DC), and by the Visitors Program of the Max-Planck 
Institute for the Physics of Complex Systems, Dresden (DC).
 

\appendix

\section{\bf Eqn. for $Q(x,s)$ and its solution for the minimal model}

From equation (\ref{main}) we get 
\begin{equation}
\frac{d^2 Q(x,s)}{d x^2}-\frac{c}{D}\frac{dQ(x,s)}{dx}-\frac{s}{D}Q(x,s)=-\frac{1}{D}\delta(x-x_{0})
\label{main1}
\end{equation}
For $x<x_{0}$, equation (\ref{main1}) has following solution.
\begin{equation}
Q(x,s)=A_{1}e^{k_{1}(L-x)}+B_{1}e^{k_{2}(L-x)}
\end{equation}
Similarly, for $x>x_{0}$, solution of equation (\ref{main1}) is,
\begin{equation}
Q(x,s)=A_{2}e^{k_{1}(L-x)}+B_{2}e^{k_{2}(L-x)}
\end{equation}
where $k_{1}$ and $k_{2}$ are given by 
\begin{equation}
k_{1}=\frac{-c-\sqrt{c^2+u^2}}{2D},  ~\rm ~  ~ k_{2}=\frac{-c+\sqrt{c^2+u^2}}{2D}
\end{equation}
with $u=\sqrt{4sD}$. 
Imposing the absorbing boundary condition at $x=0$ on $Q(x,s)$ 
we find 
\begin{equation}
B_{1}=-A_{1}e^{(k_1-k_2)L}.
\label{b1} 
\end{equation}
Now $Q(x,s)$ and it's first derivative must 
satisfy the following matching conditions at $x=x_0$:\\
\begin{equation}
A_{1} e^{k_{1}(L-x_{0})}+B_{1} e^{k_{2}(L-x_{0})}=A_{2}e^{k_{1}(L-x_{0})}+B_{2}e^{k_{2}(L-x_{0})}
\label{eq-alg1}
\end{equation}
\begin{equation}
\left.\frac{dQ(x,s)}{dx}\right |_{x=x_{0}+0}-\left.\frac{dQ(x,s)}{dx}\right |_{x=x_{0}-0}=-\frac{1}{D}
\label{eq-alg2}
\end{equation}
Reflecting boundary condition at $x=L$ gives,
\begin{equation}
A_{2} \biggl(k_{1}+\frac{c}{D}\biggr)+B_{2}\biggl(k_{2}+\frac{c}{D}\biggr)=0
\label{eq-alg3}
\end{equation}
The four unknowns $A_{1}$, $B_{1}$, $A_{2}$ and $B_{2}$ can be obtained by solving 
the three equations (\ref{b1})-(\ref{eq-alg3}). As we'll see later, 
we need $Q(x,s)$ only for $x < x_{0}$. Therefore, we give the 
expression for $A_{1}$ only in the region $x<x_{0}$: 
\begin{equation}
A_{1}=e^{\frac{V(L-x_{0})}{2D}}e^{k_{2}x_{0}}\frac{V\ \cosh\biggl(\frac{V(L-x_{0})}{2D}\biggr)-c\ \sinh\biggl(\frac{V(L-x_{0})}{2D}\biggr) }{V(V-c)e^{k_{2}L}+V(V+c)e^{k_{1}L}}
\end{equation}
with $V=\sqrt{c^2+u^2}$. And $B_{1}$ can be find by using equation \ref{b1}.
Thus, for $x<x_{0}$
\begin{equation}
Q(x,s)=A_{1}e^{k_{1}L}[e^{-k_{1}x}-e^{-k_{2}x}]
\end{equation}

where the $s$-dependence enters through the $s$-dependence of $u$ (see 
eqn.(\ref{eq-defnu}) that enters into the expression (\ref{eq-defnV}) for $V$.

\section{Model of ``Hybrid" Coupler: effects of molecular motors}
\label{sec-hybrid}

So far no motor protein has been detected at the kinetochore in budding 
yeast, which is the object of our modeling in this paper, But, in most 
of the eukaryotes (e.g., mammalian cells) such motors have been found in 
the kinetochores each of which, however, attaches to more than one MT. 
On the other hand, not more than a single MT can attach to each kt in 
budding yeat. Nevertheless, we extend our model further by including 
motor proteins \cite{sharp00b,scholey03b,salmon04}, which are active 
force generators, in addition to the passive binders in the model coupler. 
We hope to extend this hybrid coupler, in near future, by integrating 
more than one MT to model kt-MT attachments of mammalian cells. 

\begin{figure}[t]
\begin{center}
\includegraphics[width=3.45in,height=2.4in]{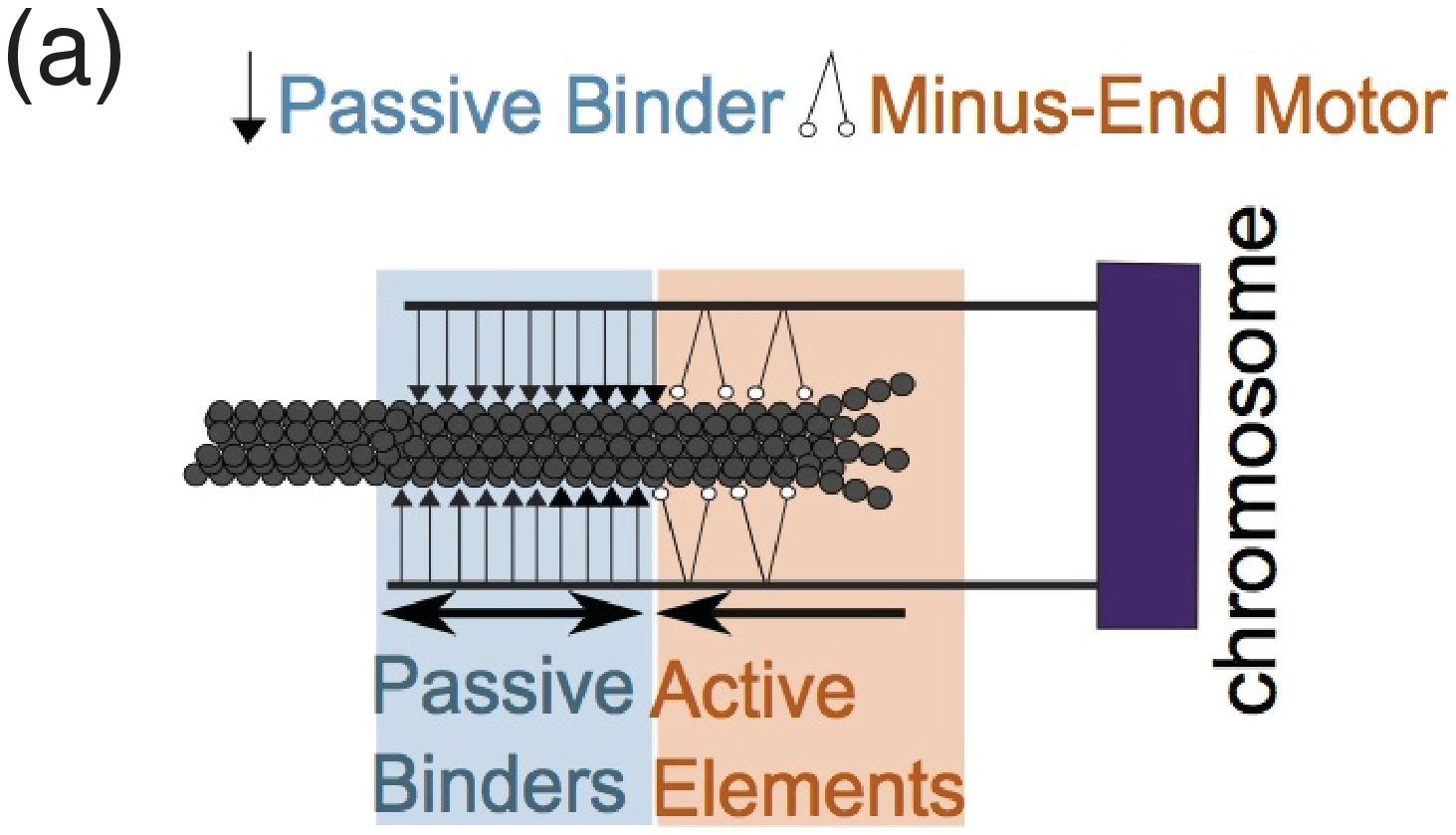}
\includegraphics[width=0.4\columnwidth]{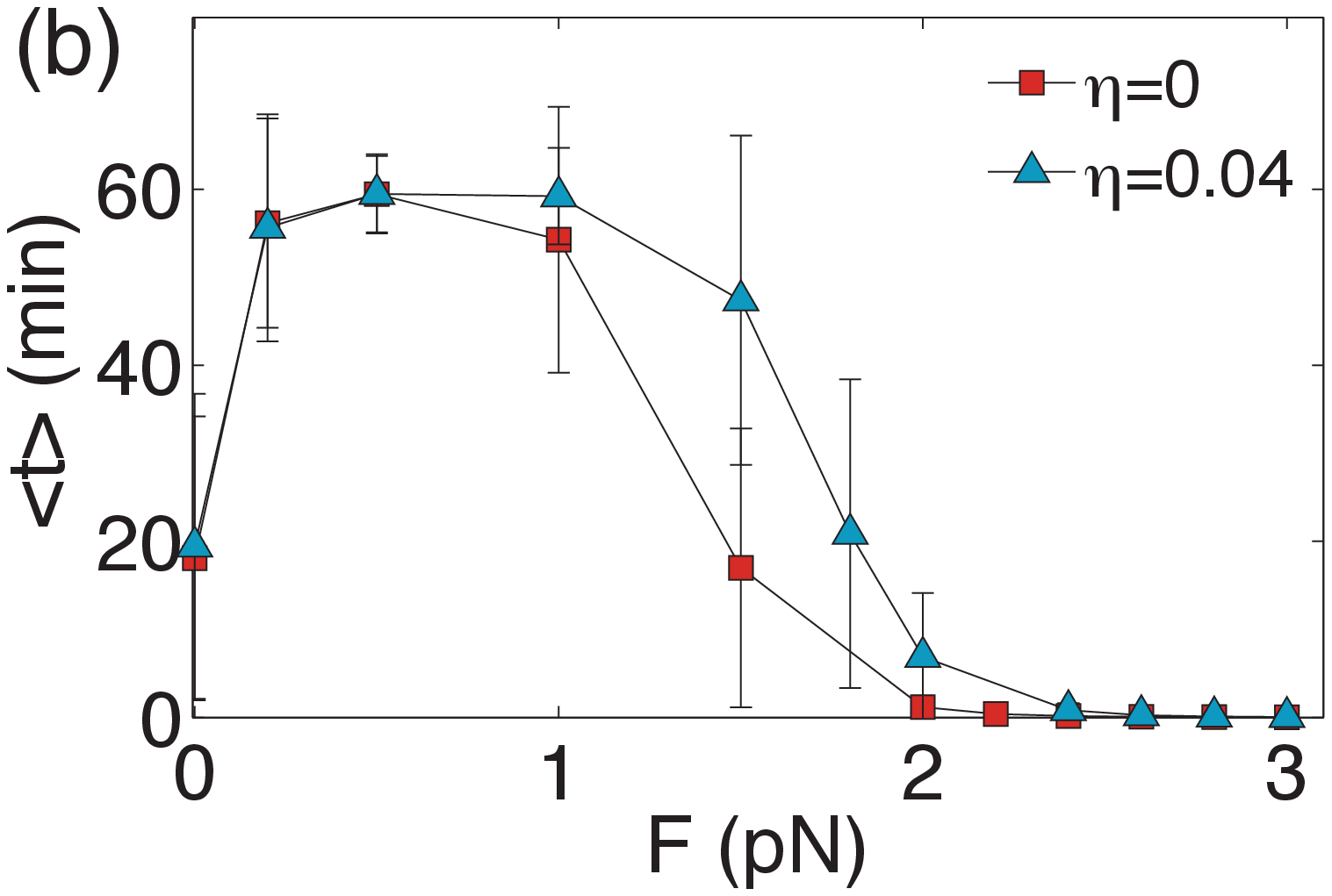}
\includegraphics[width=0.4\columnwidth]{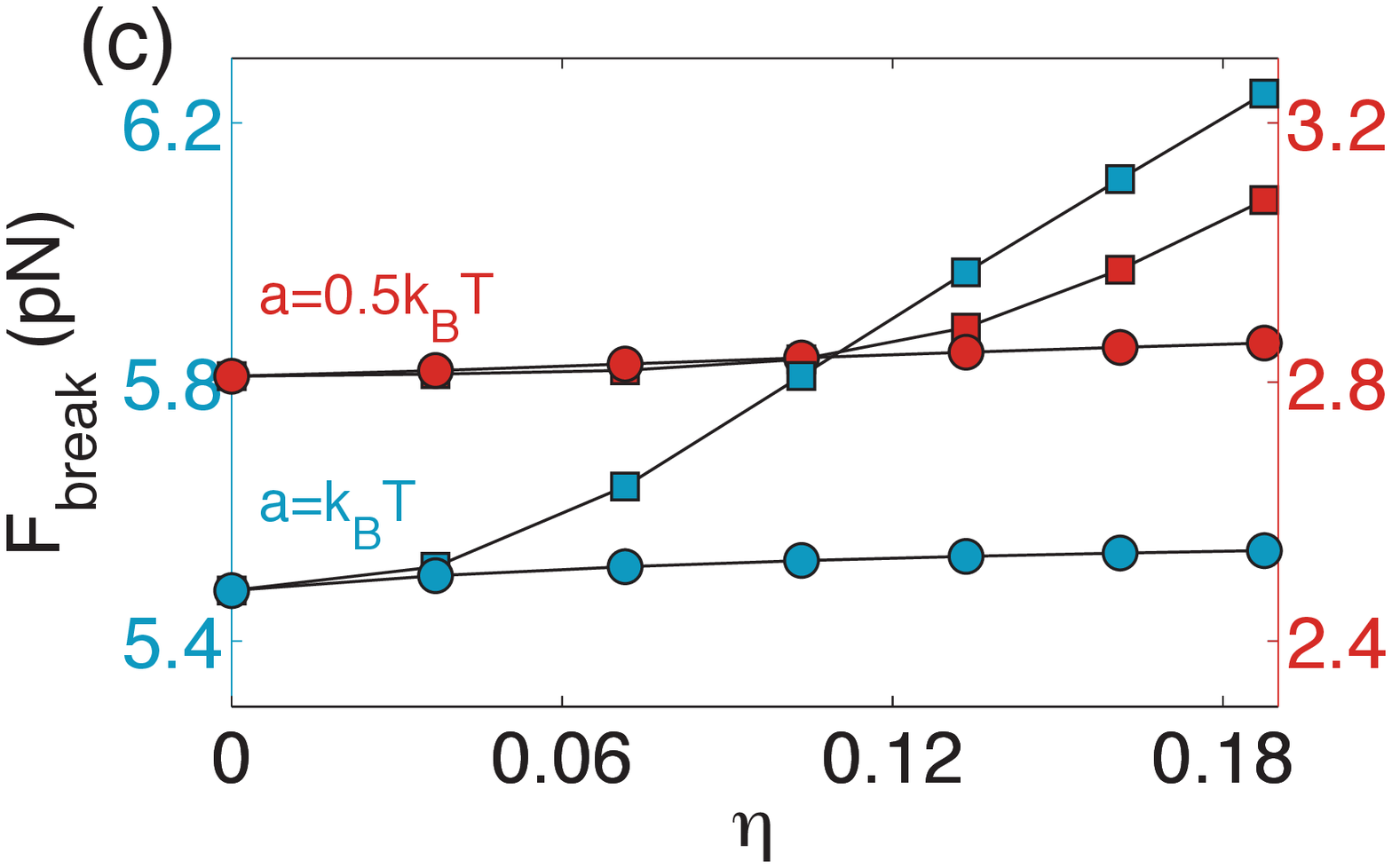}
\includegraphics[width=0.4\columnwidth]{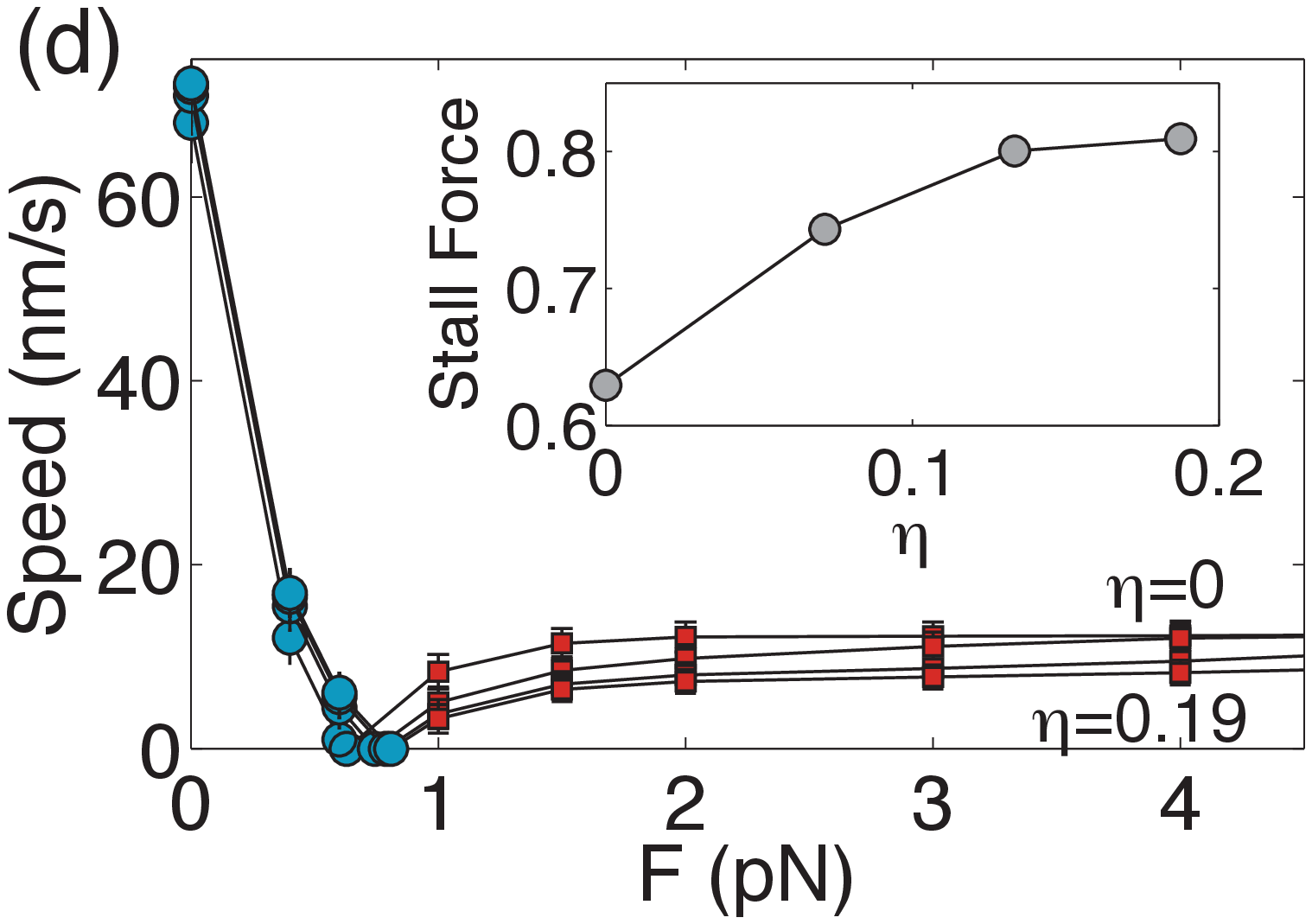}
\end{center}
\vspace{-0.2in}
\caption{ (a) Diagram of a hybrid coupler arrangement. (b) Comparison of $\langle t(L) \rangle$ for an {\it intermediate} regime hybrid coupler with ($\eta=0.04$) and without ($\eta=0$) minus-end motors with $N_{b} = 45, a = 0.45$ k$_B$T , $b = 0.04a$, $\beta_{max}= 350$ s$^{-1}$, $\alpha= 50$ s$^{-1}$. (c) Breaking load, F$_{break}$ for a low-friction hybrid coupler versus motor densities; minus-end motors (squares) and plus-end motors (circles). (d). Average speed vs force for varying densities of minus-end motors. Inset. Stall forces for each motor density fraction. Parameters for (c)-(d) are $N_{b} = 52, b = 0.01a$, $\beta_{max}= 140$ s$^{-1}$, $\alpha= 20$ s$^{-1}$. For all panels $F^{*} = 1/3$.}
\label{fig4}
\end{figure}

As indicated in recent experiments \cite{dumont12,gonen12}, the outermost 
layer of the {\it hybrid} coupler in mammalian cells consists of passive 
binders while the innermost layer, that is adjacent to the kt, is composed 
of active force generators. In the model of hybrid coupler we represent 
the active force generators by molecular motors (see Fig \ref{fig4}A).
As we show in this section, the nature of the response of the kt-MT 
attachment to the external tension now depends on whether the motors are 
plus-end directed (e.g., kinesin) or minus-end directed (e.g., dynein). 
We use the subscripts or superscripts $+$ and $-$ to refer to the plus-end 
and minus-end directed motors, respectively.

Following the general trend in the literature on molecular motors
\cite{civelekoglu06,efremov07}.
we also postulate a linear force-velocity relation for the individual motors:
\begin{align}
f_{\pm}=F_{max}^{\pm}\left( 1 - \frac{v_{\pm}}{V^{\pm}_{max}}\right),
\label{eq-fplusminus}
\end{align}
where $F_{max}^{\pm}$ and $V_{max}^{\pm}$ are the stall force and maximal velocity for the plus-end directed and minus-end directed motors, respectively, whereas $v_{\pm}$ are the corresponding instantaneous velocities.

In the dynamical equation for the overlap variable $x$, we now include 
an additional force that is generated by the motor proteins. 

We start with the force balance equation which does not include random fluctuations for the overlap velocity
\begin{equation}
 \frac{dx(t)}{dt}-V_{MT}=\frac{1}{\Gamma}\sum F=\frac{1}{\Gamma}\Big(-\Psi_{b}'(x)-F_{\text{load}}+ F_{\text{A}}(x)\Big)\label{determ_velocity},
\end{equation}
where $\Gamma$ is the effective coupler drag coefficient and $V_{MT}$ is the velocity of the MT tip with respect to a space-fixed frame of reference. The active force term
\begin{equation}
F_{\text{A}}(x)=d_{m}(x)(n_{-}f_{-}-n_{+}f_{+})
\end{equation}
with the motor density function
\begin{align}
d_{m}(x)&=(x-N_{b}{\ell})(H(x-N_{b}{\ell})-H(x-N_{b}{\ell}-L_{m}))\\
&+ L_{m} H(x-N_{b}{\ell}-L_{m}),
\end{align}
where $H(x)$ is the standard Heaviside step function and $L_{m}=8$ nm, 
corresponds to the total horizontal length of the coupler that can be populated by active components (in three-dimensions this corresponds to one layer of motors working around a MT with 13 protofilament tracks, with one motor per track).

Note that in the symbol $f_{\pm}$, the subscript minus (-) / plus (+) 
denote the force generated by the minus-end / plus-end directed motors 
that tend to increase / decrease the overlap $x$. We have expression 
(\ref{eq-fplusminus}) for $f_{\pm}$ 
where $F_{max}^{\pm}$ and $V_{max}^{\pm}$ are the stall force and maximal velocity for the plus-end directed and minus-end directed motors, respectively, whereas $v_{\pm}$ are the corresponding instantaneous velocities. 
Next we express $v_{\pm}$ in terms of $dx/dt$.  
Using $x=x_{tip}-x_{motor}$, in case of minus-end directed motors 
\begin{align}
\frac{dx}{dt}&=\frac{dx_{tip}}{dt}-\frac{dx_{motor}}{dt}\\
&=V_{MT}+v_{-}
\label{eq-vminus}
\end{align}
Similarly, if only plus-end-directed motors are present,
\begin{align}
\frac{dx}{dt}&=V_{MT}-v_{+}.
\label{eq-vplus}
\end{align}

Substituting eqs.~(\ref{eq-vminus}) and (\ref{eq-vplus}) into eq.~(\ref{determ_velocity}) we get
\begin{align}
\frac{dx}{dt}=
\frac{1}{\Gamma}&\Big[ -\Psi_{b}'(x)-F_{\text{load}} + d_{m}(x)\Big(n_{-}F_{max}^{-}\left( 1- \frac{dx/dt-V_{MT}}{V^{-}_{max}}\right)\nonumber\\
&-n_{+}F_{max}^{+}\left( 1- \frac{V_{MT}-dx/dt}{V^{+}_{max}}\right)\Big)\Big]+V_{MT}.
\end{align}

Regrouping the velocity terms we obtain the following equation for coupler overlap,

\begin{align}
\frac{dx}{dt}&=\frac{1}{\Gamma(x)}\left[ -\Psi_{b}'(x)-F_{\text{load}} + d_{m}(x)(n_{-}F_{max}^{-}-n_{+}F_{max}^{+})\right]
+V_{MT}.
\label{deteqn}
\end{align}
where $\Gamma(x)=\Gamma+\mu^{-}(x)+\mu^{+}(x)$ and
\begin{equation}
\mu^{\pm} = d_{m}(x)\biggl(\frac{n_{\pm} F^{\pm}_{max}}{ V^{\pm}_{max}}\biggr). 
\end{equation} 

The coupler movement described by eq.~(\ref{deteqn}) is fully deterministic. Next, we write down a stochastic differential equation (SDE) that would, upon averaging, correspond to the deterministic equations written above. Suppose over a small time interval $\delta t$ the number of subunits (an $\alpha-\beta$ tubulin dimer) added and removed from the tip of the MT by polymerization and depolymerization are $dN_{r}$, an independent homogenous Poisson process. We capture the effects of random Brownian forces through the noise $W(t)$ which is assumed to be a Gaussian stochastic process, which also includes the effects of fluctuations both in the chemical reactions and mechanical stepping involved in each cycle of the individual motors. Thus, finally, the equation for the hybrid coupler overlap reads
\begin{align}
dx&=\frac{1}{\Gamma(x)}\left[ -\Psi_{b}'(x)-F_{\text{load}} + d_{m}(x)(n_{-}F_{max}^{-}-n_{+}F_{max}^{+})\right]dt\nonumber\\
&+\ell dN_{r}(t)+\sigma dW(t).
\end{align}

In Fig.~\ref{fig4} we plot results for a hybrid coupler with varying density fractions of motors at the active interface, measured by $\eta=n_{\pm}/(N_{b}+n_{\pm})$. The non-monotonic nature of $\langle t(L) \rangle$ is preserved when the active components are added (see Fig.~\ref{fig4}B). If the active force generators are minus-end directed (for example, dyneins), the force exerted by them oppose load force. Consequently, the mean lifetime $\langle t(L) \rangle$ os the kt-MT attachment increases as motor fractions, $\eta$ become larger (see Figs.~\ref{fig4}C).
For a given $\eta$, the extent of such motor-induced extra stabilization of hybrid couplers can be sensitive to the binding energy of the passive components measured by $a$ (see Fig \ref{fig4}C). Despite enhancing the stability, higher numbers of load opposing motors can also lower the ability of the coupler to efficiently track rescued MT tips, as noted by the slower coupler velocities in Fig. \ref{fig4}D.
Plus-end directed motors (for example, CENP-E kinesin), which oppose overlap between the MT and the coupler, generate force that assist the external load. If the active force generators in the hybrid coupler are such plus-end directed motors, neither the coupler velocities and nor the breaking force $F_{\text{break}}$ show significant variation as $\eta$ is varied under a given tension, 
provided that there are sufficient numbers of passive binder components supporting attachment.

\end{document}